\documentclass[12pt]{article}
\newcommand{\object}{Sw 1644+57}

\usepackage{times}
\usepackage{scicite}
\usepackage{graphicx}
\usepackage{pdfpages}

\def\ale{\mathrel{\hbox{\rlap{\hbox{\lower4pt\hbox{$\sim$}}}\hbox{$<$}}}}
\def\age{\mathrel{\hbox{\rlap{\hbox{\lower4pt\hbox{$\sim$}}}\hbox{$>$}}}}

\newcommand\arcsec{\mbox{$^{\prime\prime}$}}

\newcommand{\Swift}{\textit{Swift}}
\newcommand{\event}{Sw 1644+57}

\newcommand\ion[2]{#1$\;${\scshape{#2}}}% ion, i.e., CII = \ion{C}{ii}
\def\gcn{GCN}
\def\apjl{ApJL }
\def\aj{AJ }
\def\apj{ApJ }
\def\apjl{ApJL }
\def\pasp{PASP }

\def\apjs{ApJS }
\def\araa{ARA\&A }
\def\aap{A\&A }

\def\mnras{MNRAS}

\newenvironment{sciabstract}{%
\begin{quote} \bf}
{\end{quote}}

\newcounter{lastnote}
\newenvironment{scilastnote}{%
\setcounter{lastnote}{\value{enumiv}}%
\addtocounter{lastnote}{+1}%
\begin{list}%
{\arabic{lastnote}.}
{\setlength{\leftmargin}{.22in}}
{\setlength{\labelsep}{.5em}}}
{\end{list}}

\title{An extremely luminous panchromatic outburst from 
the nucleus of a distant galaxy}

\author{\normalsize A.~J.~Levan,$^{1\ast}$ N.~R.~Tanvir,$^{2}$ S.~B.~Cenko,$^{3}$ D.~A.~Perley,$^{3}$ K.~Wiersema,$^{2}$ J.~S.~Bloom,$^{3}$ \\
\normalsize A.~S.~Fruchter,$^{4}$  A.~de Ugarte Postigo,$^{5}$ P.~T.~O'Brien,$^{2}$ N.~Butler,$^{3,6}$ A.~J.~van der Horst,$^{7}$  \\
\normalsize G.~Leloudas,$^{5}$ A.~N.~Morgan,$^{3}$ K.~Misra,$^{4}$ G.~C. Bower,$^{3}$ J.~Farihi,$^{2}$ R.~L.~Tunnicliffe,$^{1}$ \\
\normalsize  M.~Modjaz,$^{3,8}$ J.~M.~Silverman,$^{3}$  J.~Hjorth,$^{5}$ C.~Th\"one,$^{9}$ A.~Cucchiara,$^{3}$ \\  
\normalsize  J.~M.~Castro Cer\'{o}n,$^{10}$ A.J.~Castro-Tirado,$^{9}$ J.~A.~Arnold,$^{11}$ M.~Bremer,$^{12}$ J.~P.~Brodie,$^{11}$ \\
\normalsize   T.~Carroll,$^{13}$ M.~C.~Cooper,$^{14,15}$ P.~A.~Curran,$^{16}$ R.~M.~Cutri,$^{17}$ J.~Ehle,$^{13}$ , D.~Forbes,$^{18}$  \\
\normalsize J.~Fynbo,$^{5}$ J.~Gorosabel,$^{9}$ J.~Graham,$^{4,29}$  S.~Guziy,$^{9}$ D.~I.~Hoffman,$^{19}$ \\
\normalsize P.~Jakobsson,$^{20}$ A.~Kamble,$^{21}$ T.~Kerr,$^{13}$   M.~M.~Kasliwal,$^{19}$ C.~Kouveliotou,$^{22}$\\
\normalsize  D.~Kocesvki,$^{11}$ N.~M.~Law,$^{22}$ P.~E.~Nugent,$^{23}$ E.~O.~Ofek,$^{19}$ D.~Poznanski,$^{3,6,23}$ \\
\normalsize  R.~M.~Quimby,$^{19}$ E.~Rol,$^{24}$ A.~J.~Romanowsky,$^{11}$ R. S\'anchez-Ram\' irez,$^{9}$ S.~Schulze,$^{20}$ \\
\normalsize   N.~Singh,$^{11}$ R.~L.~C.~Starling,$^{2}$ R.~G.~Strom,$^{27}$ P.~J.~Wheatley,$^{1}$ R.~A.~M.~J.~Wijers,$^{26}$ \\
\normalsize J.~M.~Winters,$^{27}$ T.~Wold,$^{13}$ D.~Xu$^{28}$ \\
\normalsize{$^\ast$To whom correspondence should be addressed; E-mail:  a.j.levan@warwick.ac.uk}
}

\date{}

\begin{document} 


\baselineskip24pt


\maketitle


\begin{sciabstract}
%Transient 
Variable X-ray and $\gamma$-ray emission is characteristic of the most extreme
physical processes in the Universe, and studying the sources of these
energetic photons has been a 
major driver 
in astronomy for the past 50 years.
Here we present multiwavelength observations of a unique $\gamma$-ray
selected transient, discovered by Swift, which was accompanied by bright emission
across the electromagnetic spectrum, and whose properties are unlike any
previously observed source.  We pinpoint the event to the center of a small,
star-forming galaxy at redshift $z = 0.3534$. Its
high-energy emission 
has lasted much longer than any gamma-ray burst, while
its peak luminosity was $\sim$100 times higher than bright active
galactic nuclei.  The association of the outburst with the center of its host
galaxy suggests that this phenomenon has its origin in a new, 
rare mechanism associated with a massive black hole in the nucleus of a galaxy
\end{sciabstract}

% In setting up this template for *Science* papers, we've used both
% the \section* command and the \paragraph* command for topical
% divisions.  Which you use will of course depend on the type of paper
% you're writing.  Review Articles tend to have displayed headings, for
% which \section* is more appropriate; Research Articles, when they have
% formal topical divisions at all, tend to signal them with bold text
% that runs into the paragraph, for which \paragraph* is the right
% choice.  Either way, use the asterisk (*) modifier, as shown, to

Surveys of the sky at short wavelengths (X-ray and $\gamma$-ray) reveal a much more 
dynamic Universe than seen at optical wavelengths.
Many sources
vary substantially, and the most extreme transform from invisibility to the brightest objects in the sky, sometimes on timescales of seconds. 
The sources of such bursts of high-energy radiation have proven difficult to trace, but coherent observational 
programmes have shown that some fraction originate in the Milky Way, either from isolated
neutron stars with intense magnetic fields\cite{kouveliotou98}, or from binary systems 
containing neutron stars and black holes \cite{Aharonian05}. Some, long lived, but variable
X- and $\gamma$-ray emission originates in active galaxies \cite{Donnarumma09}, while the brightest
and perhaps most spectacular class are
the long-duration gamma-ray bursts (long-GRBs) which are detected at a rate of $\sim 2$ per week by current missions such as the {\em Swift} satellite\cite{gehrels09}, and are now thought to 
originate from the collapse of massive stars in the distant Universe \cite{hjorth03,wb2006}. 

In this paper we present observations, spanning radio to $\gamma$-ray,
of a new type of transient, 
GRB\,110328A/{\it Swift}\,J164449.3+573451, hereafter \object. It is more 
luminous than any active galaxy, 
yet longer lived than any long-GRB. It defies placement 
into any of the classes of object described above, and suggests a new channel for the creation of highly 
energetic transient events.

\object~ was first detected by the {\em Swift} Burst Alert Telescope (BAT) at 12:57:45 UT on 28 March 2011  \cite{gcn11823}. 
It was characterized as a long image trigger, with a low count rate (not sufficient to trigger the instrument) but a duration of $>1000$s, allowing
a point source to be recovered in the image plane. 
{\it Swift} follow-up 
observations with the Ultraviolet and Optical Telescope (UVOT) and X-ray Telescope (XRT) only began 1475\,s after the 
initial outburst. No source was seen in the UVOT observations, but a bright point source was found with the XRT \cite{gcn11823}. 
Unlike any previously observed long-GRB (which typically decline substantially on a timescale of minutes),
it remained bright and highly variable
for a prolonged period, and went on to re-trigger the BAT on three 
further occasions over the next 48 hours \cite{gcn11842}. Re-examination of previous $\gamma$-ray
observations of this region showed that the source appears to have been present a few days before the initial trigger, but not at earlier times \cite{gcn11891}.
Equally unlike any normal long-GRB, the source remained bright in the X-rays for more than two weeks (see Figure~1).
The early X-ray behaviour showed the same dramatic flaring seen by BAT,
with flares having time-scales of hours, and with broadly similar shapes.  
After the first 48 hours the X-rays maintained a more
constant level, albeit  with episodic brightening and fading spanning more
than an order of magnitude in flux.

\begin{figure}[ht]
\centerline{
    \includegraphics[width=16.0cm, angle=0]{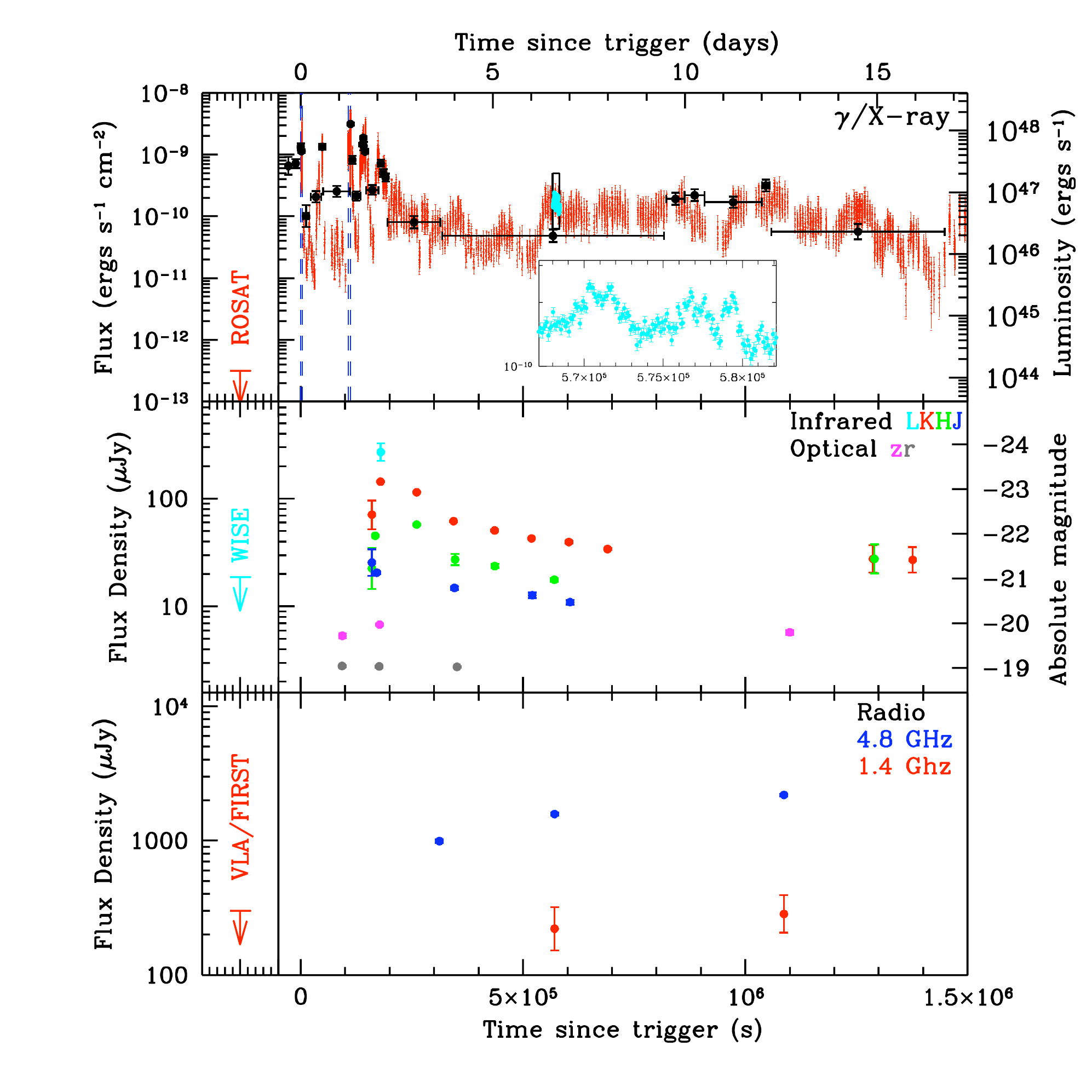}}
\caption{{\bf The X-ray, 
infrared and radio lightcurves of \object}: The top panel shows the XRT  (0.3--10\,keV; in red)
and BAT (15--50\,keV; in black)
flux against observed time since the
initial outburst trigger time, where the right hand axis indicates the luminosity of the event. 
The inset panel shows the dense sampling of our {\em Chandra} observation.
The dashed blue vertical lines show the times of subsequent triggers of the BAT. 
The middle-panel shows our
nIR lightcurve of this event (host flux not subtracted). 
The lower panel shows our 4.8\,GHz lightcurve obtained from the WSRT, demonstrating a rising radio lightcurve.  The left hand
panels represent pre-existing observation of the location of \object~and the limits on transient emission at this time \cite{bloom2011}, they 
clearly demonstrate the large amplitude of this outburst in the X-ray and infrared.}
\end{figure}

Our first ground based observations of 
\object\ began approximately two hours after the burst trigger, with the Gemini-North Telescope in Hawaii. 
Unfortunately,  poor weather conditions meant that
only shallow observations were possible, and these did not yield any 
candidate 
optical counterpart to a limit of 
$r\sim22.1$. 
At 13 hours post-trigger we obtained imaging with the Nordic Optical Telescope (NOT) on
La Palma, which revealed a $R=22.5$ magnitude source consistent with the X-ray
localisation \cite{gcn11830}.
Our examination of archival images obtained with the Palomar Transient Factory (PTF)
revealed this source to be present at approximately the same brightness
more than a year prior to the outburst, and indeed our subsequent
optical monitoring (below) confirms the optical flux is dominated
by the host galaxy. Early analysis of the X-ray/$\gamma$-ray data
was used to argue that the transient was most likely 
a source within the Milky Way \cite{gcn11824}. 
However, our spectroscopy of the optical counterpart 
with Gemini-North \cite{gcn11833}, the Gran Telescopio Canarias
(GTC)  in La Palma \cite{gcn11834}, and the 
Keck Telescope in Hawaii \cite{gcn11874}
showed strong emission lines of hydrogen and
oxygen (as well as absorption lines from a moderate age stellar population),
consistent with a star-forming galaxy
at a systemic redshift of 
$z=0.3534\pm0.0002$ (see Figure~2). 
Thus we concluded Sw 1644+57 was a source at cosmological distance
with extremely unusual properties, and this sparked a global follow-up
campaign in an effort to elucidate its nature.

We continued to monitor the field from the ground in the
optical and near infrared
(nIR) with Gemini-North, the United Kingdom Infrared Telescope (UKIRT), 
NOT, PAIRITEL  and  GTC,
obtaining observations from the $B$-band (435\,nm) to the $L$-band (3780\,nm). 
In contrast to the non-varying behaviour in the optical,
these data showed that at nIR wavelengths the source
fluctuated by more than a factor of 3 in flux over several days,
indicating that the $\gamma$-ray transient
was also producing considerable longer-wavelength emission. 
Our detection in the  $L$-band ($270\pm50\,\mu$Jy) at a level
more than an order of magnitude above our limit from 
{\it WISE} on any quiescent emission from the host galaxy at $3.4\,\mu$m
emphasises this point \cite{gcn11933}.
The infrared variations  roughly track those of the X-ray (see Figure~1), but
are certainly not perfectly correlated, suggesting multiple emission 
components.

We triggered a target of opportunity observation with the {\em Chandra X-ray Observatory}, which
took place about 6.5 days after the initial outburst.
This confirmed that the X-ray lightcurve is frequently highly variable on timescales of seconds (Figure~1).
Specifically our {\em Chandra} observations show
that factor of 2 changes in flux continued to occur on $\sim 100$ s even at comparatively late times.
However, our photometry of
individual optical and nIR images (with a time resolution of 20$-$60\,s) does not reveal rapid variability in the nIR  light. 
In the optical $r$-band
little variability was seen ($<10\%$) on all time-scales, indicating 
that the host galaxy dominates the optical emission.
We conclude that the transient has a very red optical-nIR colour,
which may be due to a high dust column along the
line of sight. The dust hypothesis would be consistent with the high hydrogen column density inferred from
the X-ray spectrum ($10^{22}$ cm$^{-2}$), which implies host extinction of $A_V \sim 6$ \cite{ps95} and together these findings suggest that
the source is situated in a dense and dusty region, such as a galactic nucleus
(see Supplementary Online Material for more information).

Observations at still longer wavelengths
showed a bright radio \cite{gcn11836}, and millimeter source \cite{gcn11841} at the same location. 
Our millimeter observations from IRAM confirm this, and radio (1.4. and 4.8 GHz) observations from Westerbork 
Synthesis Radio Telescope (WSRT) show a bright source, which brightened over the 
first week following the outburst (Figure 1, lower panel). These observations demonstrate that
\object was emitting strong radiation across the electromagnetic spectrum.

The character of the host galaxy and the position of the transient within it
are potentially important clues to the nature of \object, and
to this end, we obtained observations with the {\em Hubble Space Telescope} on 4 April 2011. 
In the nIR the image remains unresolved, consistent with 
emission from the transient still dominating, 
but in the optical wavebands we clearly detect the light of the host galaxy. 
The WFC3 IR position of the transient falls within 0.03 arcseconds ($1 \sigma$, $<150$ pc at $z=0.3534$)
of the center of the host galaxy (see Figure~3). 
We additionally obtained VLBA observations of \object~ on 1 April 2011, 
providing another precise astrometric position, with an offset from the center of the host
of $0.04 \pm 0.07$ arcseconds, further strengthening the association with the nucleus of the host
[see also \cite{gcn11911}]. 

The host galaxy itself appears compact, 
and non-interacting, 
with a half-light radius in the optical 
of $r_h = 1.04$ kpc, and an absolute magnitude of $M_V = -18.19$
(comparable in luminosity 
to the Large Magellanic Cloud). 
Subtraction of a point source from the {\em HST} F606W image 
suggests an upper limit to the transient magnitude in that band of
30\% of the host light, or a magnitude of 24.1 (AB).
The measured ratios of emission lines are consistent with them 
originating from a normal star forming galaxy that has not, 
at least until now, contained an 
active nucleus. 
The inferred star formation rate of the host is 0.5\,M$_{\odot}$\,yr$^{-1}$.

Our observations clearly show that the transient originates from the 
center of a galaxy at cosmological distances. At this redshift
(which corresponds to a luminosity distance $d_L=1.81$\,Gpc, assuming $H_0=73$\,km\,s$^{-1}$Mpc$^{-1}$,
$\Omega_{\Lambda}=0.73$, $\Omega_M=0.27$), the
the brightest X-ray flare reached a luminosity of  $L_X \sim 3 \times 10^{48}$\,erg\,s$^{-1}$ (isotropic equivalent)
for $\sim1000$\,s.
The total energy
output in the first $\sim10^6$~s after outburst of $\sim10^{53}$\,erg, 
 is equivalent to $\sim 10$\% 
of the rest energy of the Sun. While these numbers
are not abnormal
for long-duration GRBs, the properties of this outburst are clearly distinct 
from the long-GRB population. First, 
the repetition of the $\gamma$-ray trigger 4 times in 48 hours is unheard of for long-GRBs, which are destructive, and non-repeating
events. Further, the duration of bright X-ray emission is much longer than has ever been seen 
for any long-GRB [e.g., \cite{kouveliotou04,nousek06}],
persisting at
$L_X\sim10^{47}$\,erg\,s$^{-1}$ two weeks after the initial event. 
This, together with the origin in the
core of its host galaxy, lead us to
infer that \object\ most likely originates from the central massive black hole.
However, 
the X-ray luminosity of \object~ is well beyond
the bright end of the quasar luminosity function 
\cite{ueda03}, and is more luminous 
(by a factor of $\sim 100$) than flares from the brightest 
blazars [e.g., \cite{Donnarumma09}]. Interestingly, though, 
its optical luminosity is a factor of $\sim 10^4$ fainter than a bright QSO [e.g., \cite{just07}], 
implying either differing emission
processes or (as in fact seems to be the case from the red colour) a particularly
high dust column within the host. 
The overall energetics and long-duration, together with the 
order-of-magnitude variations in flux 
over 100\,s timescales, make it clear that we are observing a new and unprecedented astrophysical object (Figure~4). 

The peak luminosity  corresponds to the Eddington luminosity  of a 
$\sim 10^{10}$ M$_{\odot}$ black hole. 
It is highly unlikely that a moderate sized galaxy like the host of \object\ could contain
such a massive black hole -- indeed, our SED fitting of the host galaxy implies its total stellar 
mass is less than this value (see SOM). 
Hence \object~ is either accreting at a super-Eddington rate, or has its total energy modified by relativistic beaming (or both).
A companion paper \cite{bloom2011} considers the possibility that the source of this event is the tidal disruption of a star around the central 
black hole.

The discovery of a new class of extremely energetic
$\gamma$-ray transient after many years of intensive 
monitoring of the $\gamma$-ray sky highlights
the rarity of this phenomenon.
Studies of similar events in the future may lead to insight into the nature and fuelling of 
active galactic nuclei, or even provide
electromagnetic smoking guns of black hole -- black hole mergers, 
where the debris of the merger can result in a stellar disruption
rate of $\sim 0.1$\,yr$^{-1}$ \cite{stoneloeb}, raising the possibility that sensitive 
observations could uncover multiple events from a single galaxy. 
However, we note that
\object\ has emitted most of its radiation over the three weeks of its apparent existence relatively smoothly. It was only the presence of short, powerful bursts early on that alerted us to its presence, thus raising the possibility that other similar, but rather less variable, objects could easily be going undetected.
Future sensitive and
wide field X-ray observations of the sky offer the promise of finding more 
events like \object, as well as new classes of X-ray and $\gamma$-ray
transients.

\begin{figure}[ht]
\centerline{
    \includegraphics[width=16.0cm, angle=0]{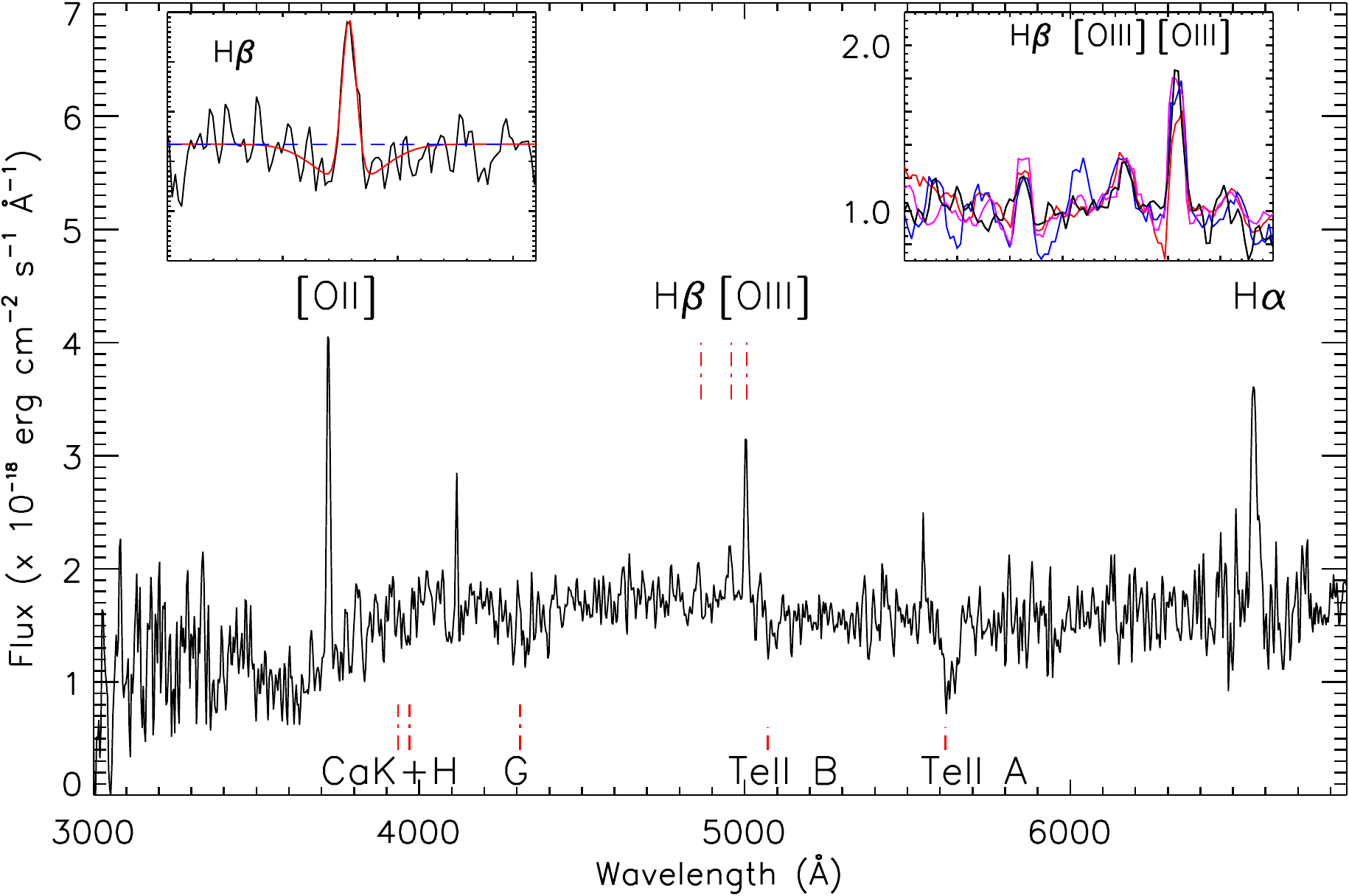}}
\caption{{\bf Spectroscopy of the host galaxy of \object}. The main panel shows our observations obtained at the GTC plotted
against rest-frame wavelength. 
The left inset shows the H\,$\beta$ line as seen in the first Gemini GMOS spectrum. Prominent stellar atmosphere absorption is visible.
The inset on the right shows the first Gemini spectrum (red), the second Gemini spectrum (blue),
the Keck spectrum (magenta) and the GTC spectrum (black) covering the H$\beta$ and [O {\sc III}] doublet, all rebinned to the lower
resolution of the GTC spectra. No emission line variability is apparent. }
\end{figure}

\begin{figure}[ht]
\centerline{
    \includegraphics[width=18.5cm, angle=0]{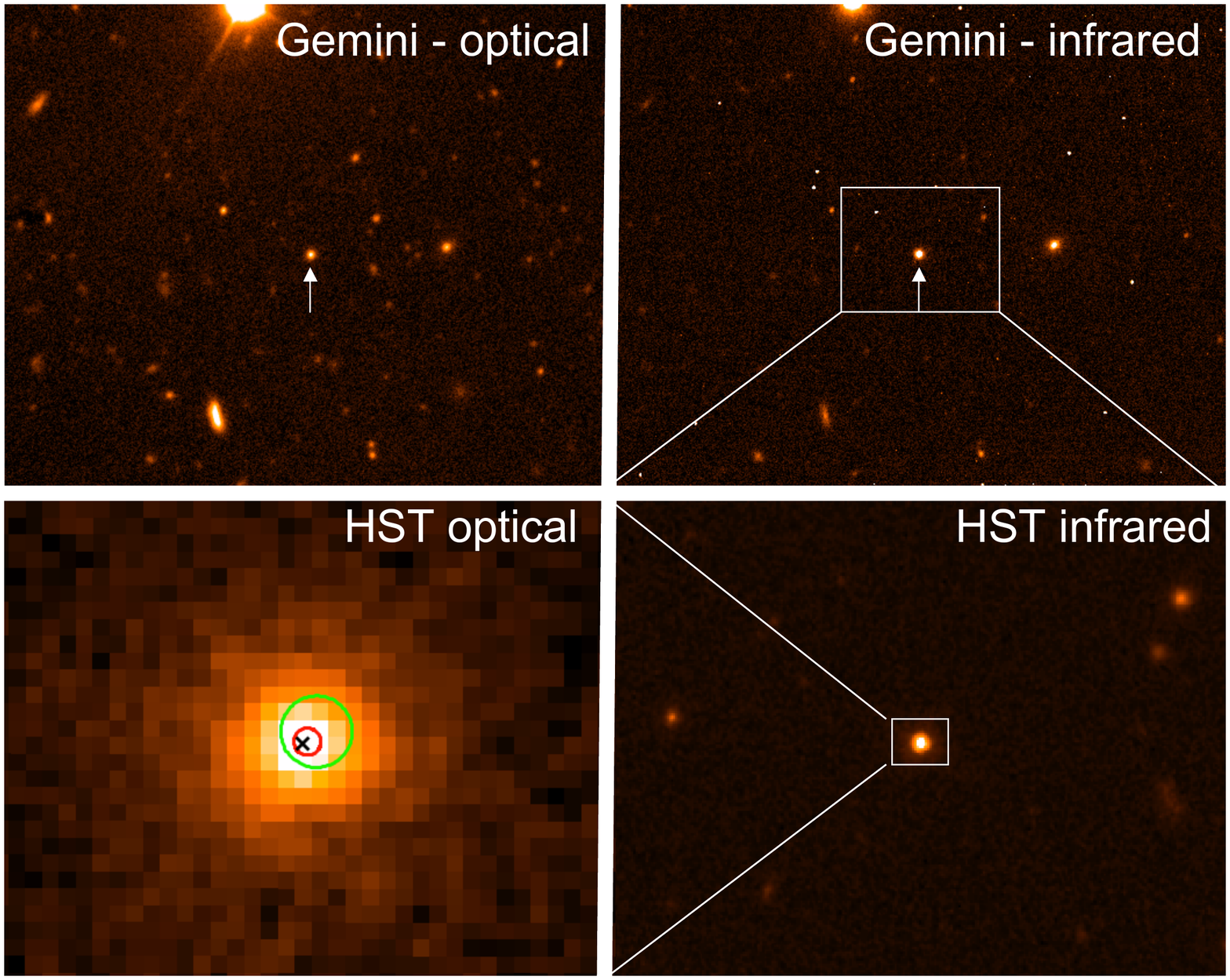}}
\caption{
{\bf Discovery images of \object~ and its host galaxy}. The top panels show our ground
based imaging in the optical $r$-band (top left), and infrared $K$-band
(top right). The images are oriented north up, east left, and are approximately
1 arcminute in height, the location of \object~ is indicated with
arrows. The lower panels show zoomed in regions of
our later
time observations with {\em HST}. The source is unresolved in our WFC IR
(F160W) imaging (bottom right), and is likely dominated by light
from the transient. In contrast the source is clearly resolved in
our WFC UVIS (F606W) imaging (bottom left, 1\,arcsecond across), and
contains at most a 30\% contribution from the transient. This image
also shows our astrometric constraints on the location of \object~
upon its host. The cross hair indicates the optically derived
centroid of the host galaxy. The blue circle shows the location of the
transient inferred from our WFC IR observations, while the larger
green circle shows the offset (due to the systematic uncertainty
in tying coordinate frames) from our VLBI position. All available
positions for the transient light are consistent with the center, and thus potentially the nucleus,
of the galaxy. 
}
\end{figure}

\begin{figure}[ht]
\centerline{
    \includegraphics[width=15.5cm, angle=0]{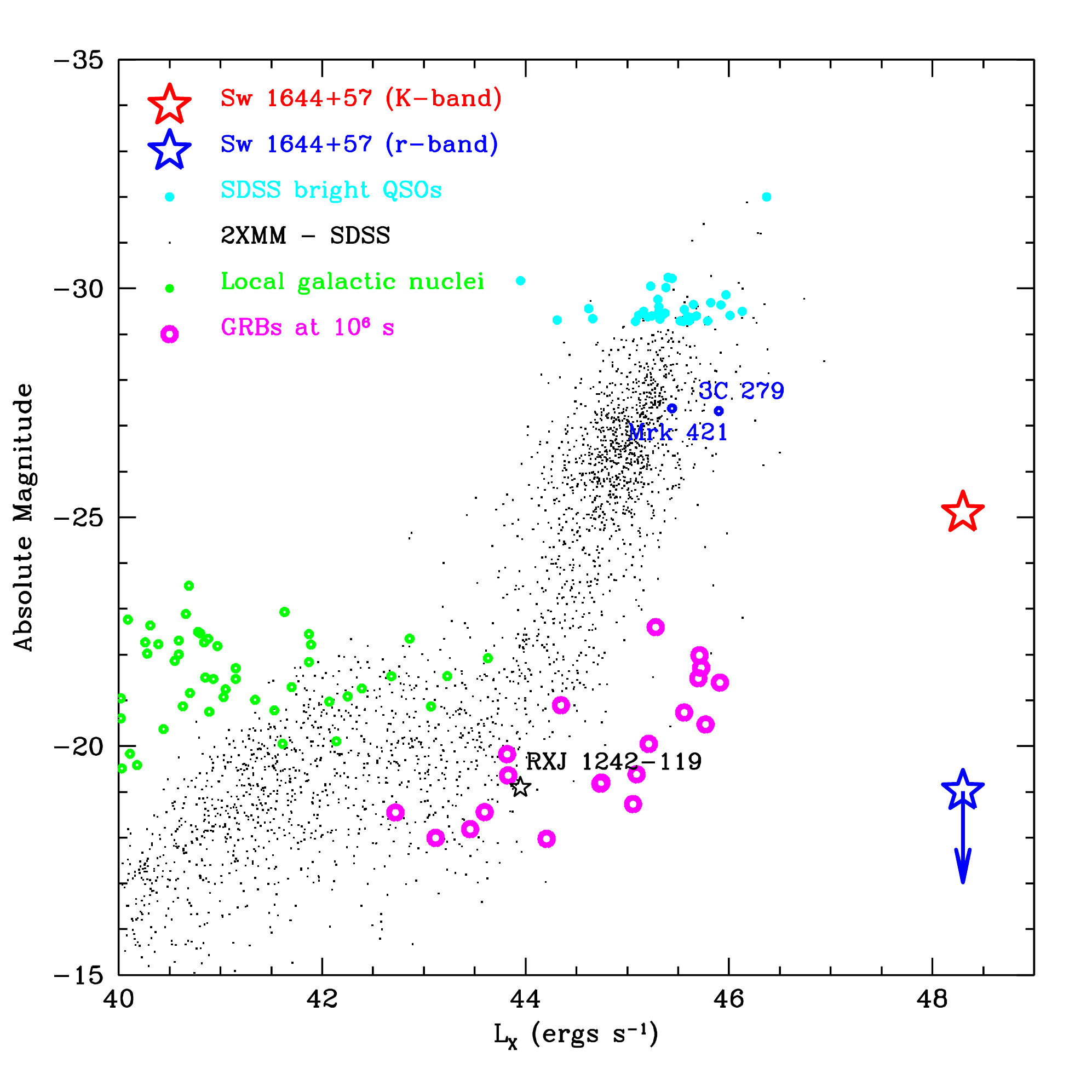}}
\caption{{\bf The uniqueness of transient \object}. The plot shows the peak X-ray luminosity and optical/nIR absolute magnitude of \object, in comparison to the 
properties of the most luminous quasars and blazars (3C 279 and Mrk 421, marked). We show for comparison a sample of all 
objects within the 2XMM survey with high confidence $(>2 \sigma)$ association with objects in SDSS of known redshift $<3$
(based on \cite{pineau11}) and a sample of more local galaxies [from \cite{ho09}, note optical magnitudes include contribution from the
host galaxy]. We also plot the late time luminosity of a sample of bright GRB afterglows
(extrapolated from \cite{kann10,evans09}), which are relevant since \object~ stays within an order of magnitude of its brightest peak,
even $10^6$ seconds after the outburst began. Finally, we also plot the location of the candidate tidal disruption event in RXJ 1242-119 
\cite{komossa2004}, It is clear that in comparison with many other objects \object~ is unique.}
\end{figure}

%\bibliography{ms}
%\bibliographystyle{Science}

\begin{scilastnote}
\item 
We gratefully acknowledge the efforts of the many observatories whose data is presented here. 
We particularly thank Daniele Malesani for assistance in the calibration
of the optical photometry, Mike Irwin for assistance with the UKIRT data, 
Derek Fox for help with the PTF data, and
Kevin Hurley and J. Xavier Prochaska for assistance in obtaining the Keck data. Full acknowledgements are given in the 
Supplementary online material. 

\end{scilastnote}

\author{
\footnotesize{$^{1}$Department of Physics, University of Warwick, Coventry, CV4 7AL, UK}\\
\footnotesize{$^{2}$Department of Physics and Astronomy, University of Leicester, Leicester, LE1 7RH, UK}\\
\footnotesize{$^{3}$Department of Astronomy, University of California, Berkeley, CA 94720-3411, USA}\\
\footnotesize{$^{4}$Space Telescope Science Institute, 3700 San Martin Drive, Baltimore, MD 21218, USA}\\
\footnotesize{$^{5}$Dark Cosmology Centre, Niels Bohr Institute, University of Copenhagen, 2100 Copenhagen, Denmark}\\
\footnotesize{$^{6}$NASA Einstein Fellow}\\
\footnotesize{$^{7}$Universities Space Research Association, NSSTC,
320 Sparkman Drive, Huntsville, AL 35805, USA}\\
\footnotesize{$^{8}$Columbia Astrophysics Lab, Columbia University, NYC, NY 10024, USA}\\
\footnotesize{$^{9}$Instituto de Astrof\' isica de Andaluc\' ia (IAA-CSIC), Glorieta de la Astronom\' ia s/n, E-18008 Granada, Spain.}
\footnotesize{$^{10}$Herschel Science Operations Centre, ESAC, ESA, PO Box 78, 28691 Villanueva de la Ca–ada, Madrid, Spain }\\
\footnotesize{$^{11}$UCO/Lick Observatory, University of California, Santa Cruz, 1156 High Street, Santa Cruz, CA 95064, USA}\\
\footnotesize{$^{12}$Institut de RadioAstronomie Millim\'etrique, 300 rue de la Piscine, Domaine Universitaire,
38406 Saint Martin d'H\`eres, France}\\
\footnotesize{$^{13}$Joint Astronomy center, 660 North A'ohoku Place, University Park, Hilo, HI 96720, USA}\\
\footnotesize{$^{14}$Center for Galaxy Evolution, University of California, Irvine, 4129 Frederick Reines Hall, Irvine, CA 92697}\\
\footnotesize{$^{15}$Hubble Fellow}\\
\footnotesize{$^{16}$ AIM, CEA/DSM - CNRS, Irfu/SAP, Centre de Saclay, Bat.
709, FR-91191 Gif-sur-Yvette Cedex, France }\\
\footnotesize{$^{17}$Infrared Processing and Analysis Center, California Institute of Technology, Pasadena, CA, 91125, USA}\\
\footnotesize{$^{16}$Centre for Astrophysics \& Supercomputing, Swinburne University, Hawthorn VIC 3122, Australia}\\
\footnotesize{$^{18}$Institute for the Physics and Mathematics of the Universe, University of Tokyo, Kashiwa-shi, Chiba 277-8568, Japan}\\
\footnotesize{$^{19}$Cahill Center for Astrophysics, California Institute of Technology, Pasadena, CA, 91125, USA}\\
\footnotesize{$^{20}$Centre for Astrophysics \& Cosmology, Science Institute, University of Iceland, Dunhaga 5, IS-107 Reykjavik, Iceland}\\
\footnotesize{$^{21}$Center for Gravitation and Cosmology, University of Wisconsin-Milwaukee, 1900 E Kenwood Blvd, Milwaukee, WI - 53211, USA}\\
\footnotesize{$^{22}$Space Science Office, VP62, NASA/Marshall Space Flight Center Huntsville, AL 35812, USA}\\
\footnotesize{$^{23}$SLAC National Accelerator Center, Kavli Institute for Particle Astrophysics and Cosmology, 2575 Sand Hill Rd, MS 29, Menlo Park, Ca 94025, USA}\\
\footnotesize{$^{24}$Dunlap Institute for Astronomy \& Astrophysics, University of Toronto, Toronto M5S 3H4, Ontario, Canada}\\
\footnotesize{$^{25}$Computational Cosmology Center, Lawrence Berkeley National Laboratory, 1 Cyclotron Road, Berkeley, CA 94720, USA}\\
\footnotesize{$^{26}$Astronomical Institute, University of Amsterdam, Science Park 904, 1098 XH Amsterdam, The Netherlands}\\
\footnotesize{$^{27}$Netherlands Institute for Radio Astronomy (ASTRON),
Postbus 2, 7990 AA Dwingeloo,  The Netherlands}\\
\footnotesize{$^{28}$Benoziyo Center for Astrophysics, Faculty of Physics, Weizmann Institute of Science, Rehovot, 76100, Israel} \\
\footnotesize{$^{29}$Department of Physics and Astronomy, Johns Hopkins University, Baltimore, MD 21218}\\
}

\clearpage

\begin{center}
{\Huge {\bf Supplemental Online Material}}:  \\ {\LARGE ``An extremely luminous panchromatic outburst from 
the nucleus of a distant galaxy"}
\end{center}

\footnotesize{
We adopt cosmological parameters of 
$H_0=73$\,km\,s$^{-1}$Mpc$^{-1}$, $\Omega_\Lambda = 0.73$ and   $\Omega_m = 0.27$. 
At a redshift of $z=0.3534$, the luminosity distance is 1814.8 Mpc and 1 arcsec represents 4.803 kpc in projection. These
are identical parameters to those used in our companion paper \cite{bloom2011}

\section{The uniqueness of \event}
In the main text we argue (in particular in Fig.~4) that \event\ does
not belong to any known class of high-energy transient sources, and
therefore represents a novel astrophysical phenomenon.  Here, we
provide further justification of this claim by comparing \event\ to 
other luminous optical, X- and $\gamma$-ray sources, in particular 
gamma-ray bursts (GRBs) and active galactic nuclei (AGN). 

\subsection{Comparison with gamma-ray bursts}
A relatively wide variety of astrophysical sources are capable of
generating sufficiently bright outbursts of high-energy radiation to
trigger the current generation of gamma-ray satellites.  These sources
span an incredible range of the observable universe, from electrical
discharges associated with thunderstorms on Earth \cite{fbm+94} to
the deaths of the earliest known stars \cite{tfl+09,sdc+09}.

The association of \event\ with a galaxy at $z = 0.3534$ 
immediately rules out all but the most luminous of these
gamma-ray outbursts.  While repeated high-energy triggers and rapid
X-ray variability at late times have been seen before in events
discovered by the \Swift\ satellite\footnote{For example, soft 
gamma-ray repeaters \cite{wt06} can generate multiple high-energy
triggers, and the unusual Galactic transient GRB\,070610 exhibited
dramatic X-ray and optical flaring not typically seen from gamma-ray
bursts \cite{kck+08,sks+08,cdg+08}.}, these sources are not known to
generate anywhere near the total energy release observed from \event\
($E_{\mathrm{iso}} \approx 10^{53}$\,erg).  

Gamma-ray bursts (GRBs) are the most luminous class of these
high-energy transients ($L_{\mathrm{iso}} \approx
10^{50}$--$10^{52}$\,erg\,s$^{-1}$), and \event\ was initially
classified as a long-duration [i.e., resulting from massive star
core-collapse \cite{wb06}] GRB \cite{gcn11823}.  But two important
properties distinguish \event\ from long-duration GRBs\footnote{Given
  the extreme duration and large luminosity, we also consider it
  extremely unlikely \event\ is a member of the short-duration class
  of GRBs.}, clearly
establishing it as a distinct class of high-energy transient.

First, unlike soft gamma-ray repeaters, long-duration GRBs are
destructive events.  The duration of the prompt gamma-ray emission,
dictated by stellar debris accreting onto the newly-formed compact
object, is typically $\Delta t \sim 1$--10\,s.  While late-time ($t
\gg \Delta t$) engine activity can manifest as bright X-ray flares
\cite{burrows05}, no long-duration GRB has ever caused subsequent
gamma-ray triggers of the \Swift\ satellite in the manner of \event.
The dramatically different timescale associated with the ``prompt''
emission (4 triggers over a period of 48\,hours) is strongly
suggestive of a distinct origin from long-duration GRBs.

In addition to the longer time scale associated with the gamma-rays, 
the X-ray (0.3--10\,keV) emission from \event\ is quite unlike any
known GRB afterglow.  In Figure~\ref{xrtcomp} we compare
the X-ray light curve of \event\ with a representative sample of
long-duration GRBs.  While many GRBs exhibit an extended period of
relatively flat emission [the so-called ``plateau'' phase
\cite{nkg+06}], no other long-duration GRB declines so little over the
first $\sim 2$\,weeks of its evolution.  The long-lived light curve
also makes the late-time emission significantly more luminous than
normal long-duration GRBs.  Together, the longer time
scales associated with both the X-ray and gamma-ray emission, as well
as the large X-ray luminosity, are indicative that a significantly
more massive black hole is responsible for the observed emission 
from \event\ [an idea explored in detail in our companion paper \cite{bloom2011}].

Finally, we note that the astrometric coincidence with the center of
the host galaxy is relatively unique amongst long-duration GRBs (at
least those with observed with sufficiently high angular resolution).
Given their association with star formation, it is not surprising that
in general the location of long-duration GRBs is highly correlated
with host galaxy light (in particular the blue light generated by
massive stars \cite{bkd02,fls+06}).  It is nonetheless unusual (though
not unprecedented) for a long-duration GRB afterglow observed with 
the exquisite angular resolution of \textit{HST} to be astrometrically
consistent with the nucleus of its host galaxy.

\begin{figure}[ht]
\centerline{
%	\centerline{\psfig{file=complin.ps,angle=270,width=9cm,clip=5mm}}
    \includegraphics[width=7.5cm, angle=0]{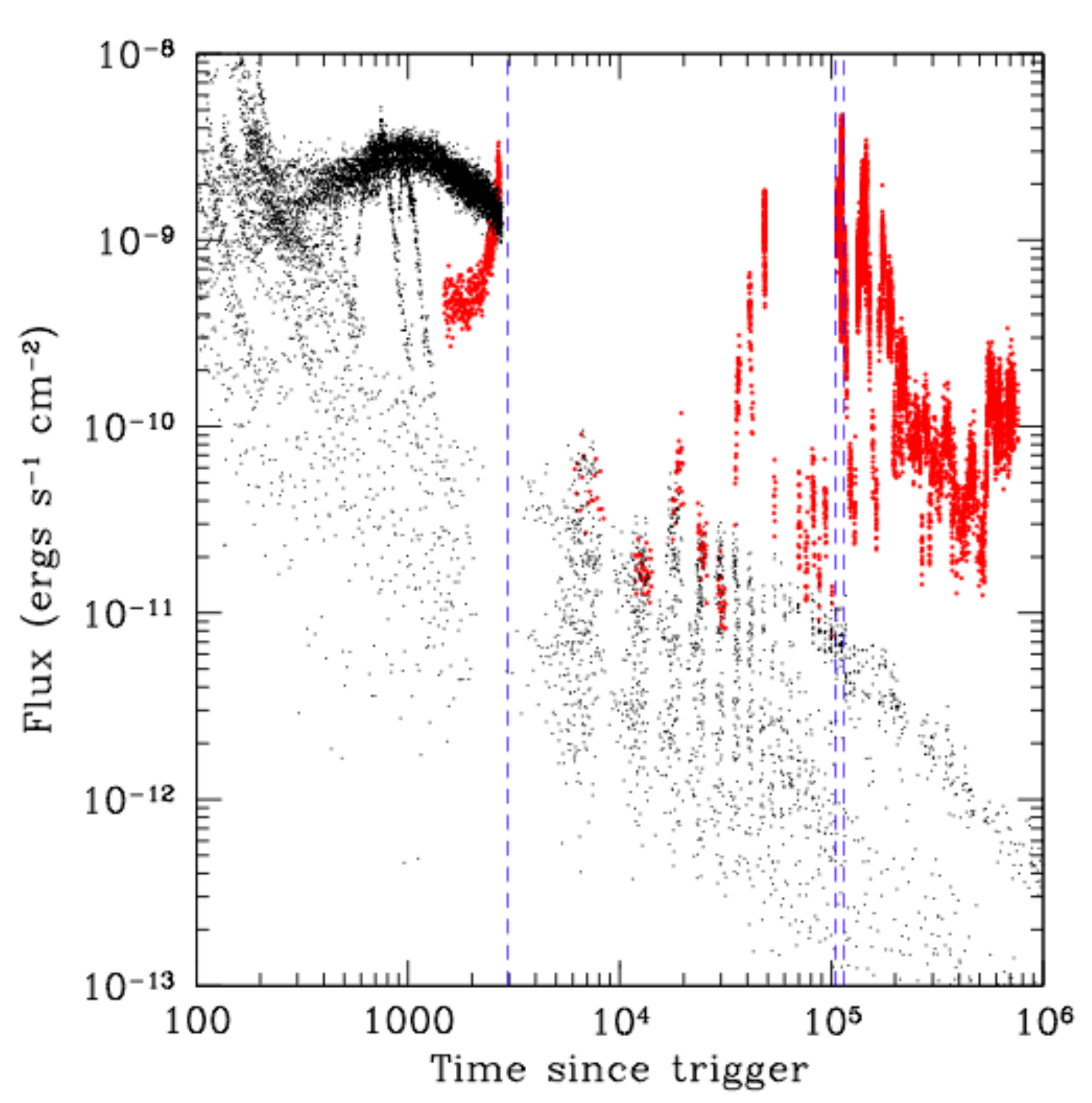}
    \includegraphics[width=7.5cm, angle=0]{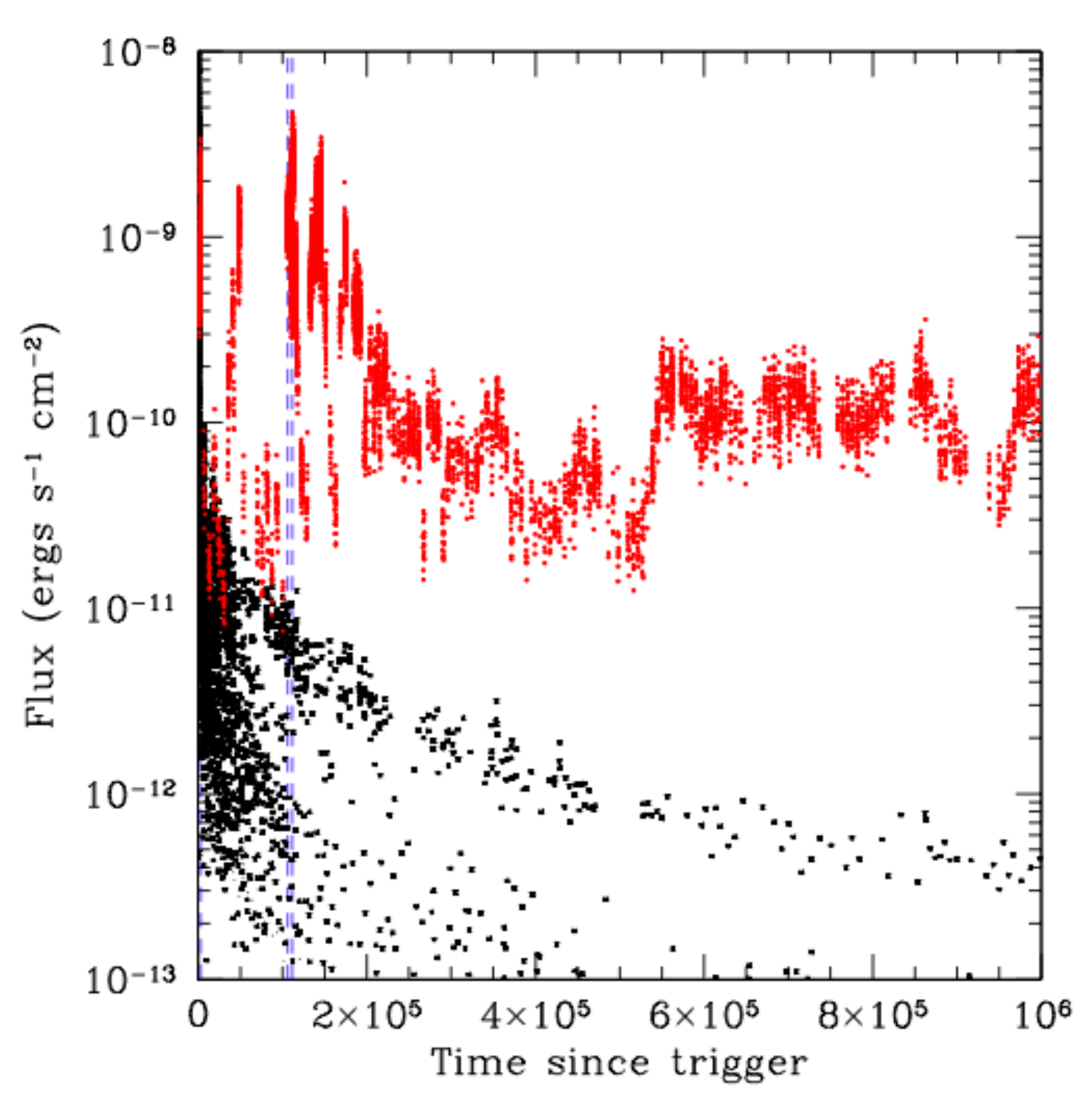}}
\caption{A comparison of the properties of \event~ with those of GRB afterglows [from \cite{evans09}].  The left hand panel shows a logarithmic time and flux axis, since
GRBs usually decay as powerlaw's. The right hand panel shows the same plot, but with a linear time axis. It is clear that \event\ is much longer lived
than any GRB engine yet observed, and is two orders of magnitude brighter (in flux) at $10^6$\,s than observed for the brightest GRBs. No other GRB at similar
redshift has ever appeared so bright at late times, and the afterglow of \event~ is also a factor of 10-100 brighter in luminosity at late times than the
brightest GRBs (see Figure 4, main article). }
\label{xrtcomp}
\end{figure}

\subsection{Comparison with active galactic nuclei} 
Given the long lived nature of the emission, and its location in the core of the host galaxy, it is also relevant to consider the comparison to the different manifestations
of  active galactic nuclei (AGN). The majority of AGN (e.g. LINERS, Seyferts, etc.) 
are of moderate X-ray and optical luminosity, and in this respect are not
%are clearly not 
%of interest to explain 
comparable to an exceptionally bright
object like \event. 
The brighter AGN are the luminous quasars, and the blazars. These can reach X-ray luminosities approaching $10^{46}$ erg s$^{-1}$, with
corresponding optical absolute magnitudes of $-30$ or brighter. 
Furthermore, the blazars are intrinsically the most variable AGN, and can show rapid 
$\sim 2$ factor variation in their lightcurves. However, \event\ exhibits much more rapid variability, of order a factor of 100 in just a few hundred seconds,
and is also reaches peak luminosity in excess of $10^{48}$ ergs s$^{-1}$. 
This is shown graphically in Figure~4 of the main text where we compare the X-ray luminosity and optical absolute magnitude of \event~ against those
of QSO, Blazars, other AGN and GRB afterglows 10$^6$ seconds post burst.  As can be seen, \event~ stands 
%well distinct 
apart from any object previously observed,
and leads us to conclude we are dealing with a previously unobserved, and extremely 
%luminous 
energetic phenomenon.

\section{Observations}

\subsection{Pre-event imaging}
\subsubsection{Palomar Transient Factory (PTF)}

As part of the Palomar Transient Factory [PTF
\cite{lkd+09,rkl+09}], 
%we have obtained 
pre-outburst optical 
($R$-band) imaging of the field of \event\ 
were obtained with the Palomar 48\,inch 
Oschin Schmidt telescope over the time period from 2009 May to
2010 October.  In a stacked frame of all available data (77 individual 
images), we find a faint, unresolved source at location $\alpha = 
16^{\mathrm{h}} 44^{\mathrm{m}} 49.983^{\mathrm{s}}$, $\delta =
+57^{\circ} 34' 59.72'$ (J2000.0).  Astrometry was performed 
relative to 23 objects from the 2MASS point source catalog 
\cite{2MASS}.  The uncertainty in this position, including both the 
statistical error in the centroiding process and systematic error from 
the astrometric tie, is $\approx 250$\,mas in each coordinate.  Using 
stars in the field calibrated via our LRIS imaging, we measure a 
magnitude of $R = 22.09 \pm 0.26$.  This is marginally brighter than
the value derived from our post-outburst imaging; however, given the 
somewhat different filters and the large uncertainty in this
measurement, we do not consider this discrepancy particularly significant.

\subsubsection{Wide-field Infrared Survey Explorer (WISE)}

In addition to the optical imaging described above, the location of \event~ was
observed by the Wide-field Infrared Survey Satellite (WISE) \cite{wise} between 
25 - 28 Jan 2010. Observations were obtained in each of 
the 3.4, 4.6, 12 and 22 micron bands (W1--4 respectively), with
a total of 39 individual images per band included in each filter co-add. 
The resulting limits are shown in Table~\ref{wise}. These images
suggest the host galaxy itself is not extremely red, and confirm
that the vast majority of the flux in our mid-IR measurement
(see below) is originating in the afterglow.

\begin{table}[htdp]
\caption{Catalog of WISE observations}
\begin{center}
\begin{tabular}{lll}
\hline
Band & Limit (mag) & Flux ($\mu$ Jy) \\
\hline
W1 & $>18.0$ & $>19$ \\
W2 & $>16.2$ & $>57$ \\
W3 & $>12.8$ & $>240$ \\
W4 & $>9.55$ & $>1270$ \\
\hline
\end{tabular}
\end{center}
\label{wise}
\end{table}%

\subsection{Optical imaging}

%Burst MJD is 55648.54010417
\begin{table}
\begin{center}
\begin{tabular}{llllll}
\hline
Start (MJD)             &        Mid-point (s, post burst)         & Telescope     & Filter & exptime (s) & Magnitude \\
\hline
55648.6466644 & 9207 &                                  Gemini-N        & $r$ & 240 & $>22.40$ \\
55649.6094537   &       92797                           &       Gemini-N        & $r$ & $5 \times 120$ & 22.78 $\pm$ 0.02   \\
55649.5994062 &         99273                           &     Gemini-N  & $g$ &  $5 \times 120$  & 23.66 $\pm$ 0.05   \\
55649.6194815 &        93662                            &     Gemini-N       & $z$ &   $5 \times 120$   &  22.08 $\pm$ 0.05  \\
55650.5779583 &      176475                       &       Gemini-N       & $r$ & $5 \times 120$& 22.79 $\pm$ 0.02 \\
55650.5879664  &        177340                     &      Gemini-N         & $z$  & $5 \times 120$ & 21.82 $\pm$ 0.02  \\
55652.6031910 & 	351462				&	Gemini-N		& $r$ & $5 \times 120$ & 22.80 $\pm$ 0.02 \\
55649.1858626 &	57121		&		GTC 	& $i$ & $4 \times 60$ & 22.29 $\pm$ 0.05 \\
55661.2350123 & 	1100079	&		GTC 	& $z$ & $8 \times 90$ & 22.00 $\pm$ 0.05 \\
55655.56435 & 607871		&		Keck & B & $5 \times 330$  & 24.35 $\pm$ 0.05 \\
55655.56480 &	 607942	&		Keck & i  & $5 \times 300$ & 22.31 $\pm$ 0.02 \\
55649.0744306 & 50036 & NOT & $B$ & 2 $\times$ 600 &24.40  $\pm$ 0.10 \\
%55649.0842616 & 48208 & NOT  & $z$ & $4 \times 300$& \\
 55649.1199514 & 50745  & NOT & $V$ & $2 \times 600$& 23.17 $\pm$ 0.05 \\
55649.1361632  & 52146 & NOT & $R$ & $2 \times 600$& 22.55 $\pm$ 0.03 \\
%55650.1737442 & 142689 & NOT & $z$  & $8 \times 300$& \\
55650.2109861 & 145362 & NOT & $R$ & $3 \times 600$& 22.51 $\pm$  0.04 \\
55656.1730903 & 660436 & NOT & $R$  & $3 \times 600$& 22.40 $\pm$  0.06 \\
55661.0701887 & 1083334 & NOT & $R$ & $4 \times 300$ & 22.44 $\pm$ 0.06 \\
\hline
\end{tabular}
\end{center}
\caption{Optical photometry of \event, obtained from Gemini-N, the GTC and the NOT. }
\label{opdata}

\end{table}%

We obtained extensive optical imaging of \event~ using the Gemini-North Telescope, the Nordic Optical Telescope (NOT), the Gran Telescopio CANARIAS (GTC)
and the Keck I telescope. A log of optical observations is shown in Table~\ref{opdata}. The data were all processed through IRAF in the normal way, and
magnitudes and fluxes were extracted via aperture photometry of the source in comparison to objects in the field. Photometric calibration was based
on standard stars, observed by the Nordic Optical Telescope, and confirmed by comparison to published zeropoints for our Gemini-North observations. 

To search for variability we performed point spread function matched image subtraction, using the ISIS package \cite{alard2000}. To obtain the
cleanest subtractions we limit ourselves to 
images taken from the same telescope on different nights. Since each telescope
offers a reasonable time baseline this enables us to search for optical variability 
over the course of the first week post burst. 

In our optical ($r$- and $R$-band) imaging there is no evidence for any variable source within the host galaxy of \event,
with the most constraining limits coming from our Gemini observations. 
To estimate the limiting magnitude of these subtractions
we seed our first epoch images with artificial point spread functions of known magnitude, and use these to photometrically calibrate our subtracted images. 
From this we estimate that any variable source within the host of \event\ had a magnitude fainter than $r>25.5$ ($3\sigma$), in our earliest observations. 
This evidence does not necessarily preclude the presence of an optically variable 
transient which contributes similar flux to all epochs over our
range of observation to date, however,
it suggests that the variation in the flux between the observations was $<0.23$ $\mu$Jy. 

While our $r$-band observations do not yield any evidence of transient emission, we clearly detect a variable source in our $z$-band imaging, as shown in Figure~\ref{psfsub}. 
The source is seen to brighten between our first and second epoch of $z$-band observations which were taken 26 and 49 hours after the initial outburst. The X-ray
lightcurve over the same time period is also seen to brighten, although by a much larger amplitude.

\begin{figure}[ht]
\centerline{
%	\centerline{\psfig{file=complin.ps,angle=270,width=9cm,clip=5mm}}
    \includegraphics[width=15cm, angle=0]{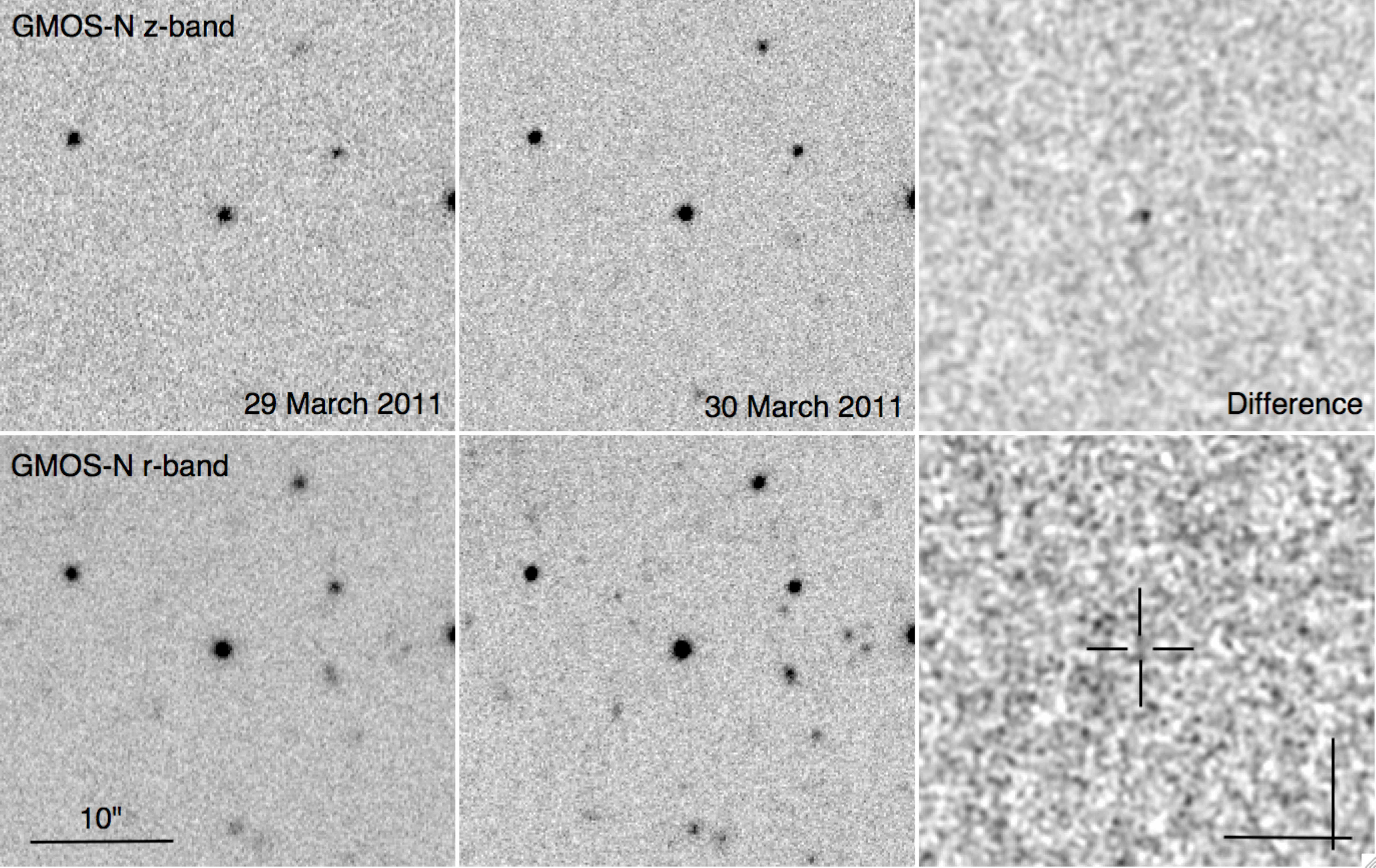}}
\caption{Our Gemini/GMOS imaging of \event. The top panels show of $z$-band imaging obtained on 29 and 30 March 2011, and the
result of a PSF matched image subtraction. The bottom panels represent the same, but for our $r$-band imaging. In the $z$-band we
clearly see residual emission, brightening between the two observations. In the $r$-band, no such variation is obvious, consistent
with \event\ originating on a highly dusty sightline which diminishes
the transient optical flux.  Images are North up and East to the left }
\label{psfsub}
\end{figure}

\subsection{Infrared imaging}

We obtained IR observations of \event with PAIRITEL, UKIRT and Gemini-North. A log of observations is
shown in Table~\ref{ir}. The PAIRITEL and UKIRT observations were both pipeline processed using dedicated, instrument specific pipelines. 
The Gemini-North data were reduced using the standard NIRI package within IRAF. Photometric calibration was taken relative
to 2MASS, and the resulting photometry is shown in Table~\ref{ir}. 

We additionally obtained thermal IR observations of \event~ with NIRI in the $L$-band. 
For this we used the 30 co-adds of 1\,s exposures at
each of 58 random dither positions, and used the f/32 camera to avoid saturation. 
Unfortunately in this imaging both the science and
standard observations (of FS 138) showed substantial elongation of the point spread 
function. However, we were able to extract  detection
of the afterglow of \event. This provides a magnitude of $L'=14.9 \pm 0.2$ (vegamag), or, 
correspondingly a flux of $0.27 \pm 0.05$\,mJy at 3.78\,microns. 

%Burst MJD is 55648.54010417

\begin{table}
\begin{center}
\begin{tabular}{lllll}
\hline
MJD start & Mid point & Mag & Telescope & Band \\
\hline
55649.35609 & 71588 &$>19.1$ & PAIRITEL & J \\
55650.34888 & 161627 & 19.511 $\pm$ 0.284 & PAIRITEL & J \\
55650.49898 & 170821 &19.747 $\pm$ 0.023 & UKIRT & J \\
55652.52602 & 345289 &20.100 $\pm$ 0.040 & UKIRT & J \\
55654.55962 & 520974 &20.270 $\pm$ 0.057 & UKIRT & J \\
55655.53639 & 605811&20.429 $\pm$ 0.049 & UKIRT & J \\
\hline
55649.35609 & 71588 &$>18.2$ & PAIRITEL & H \\
55650.34888 & 161627 & 19.068 $\pm$ 0.435 & PAIRITEL & H \\
55650.45965 & 167400 &18.311 $\pm$ 0.023 & UKIRT & H \\
55651.55957 & 261049 &18.050 $\pm$ 0.019 & Gemini & H \\
55652.55021 & 347371 &18.861 $\pm$ 0.023 &UKIRT & H \\
55653.58889 & 436380 &19.014 $\pm$ 0.044 & Gemini & H \\
55655.13265 & 570064 &19.330 $\pm$ 0.040 & HST & H \\
55663.47546 & 1290643 & 18.846 $\pm$ 0.309 & UKIRT & H \\
\hline
55649.35609 & 71588 & $>16.9$ & PAIRITEL & K \\
55650.34888 & 161627 & 17.353 $\pm$ 0.306 & PAIRITEL & K \\
55650.60548 & 178995 &16.586 $\pm$0.007 & Gemini & K \\
55651.55476 & 260634 &16.833 $\pm$ 0.010 & Gemini & K \\
55652.50449 & 343327 &17.502 $\pm$ 0.015 & UKIRT & K \\
55653.58405 & 435965 &17.717 $\pm$ 0.019 & Gemini & K \\
55654.53582 & 518918 &17.903 $\pm$ 0.022 & UKIRT & K \\
55655.50653 &603003 &17.988 $\pm$ 0.022 & UKIRT & K \\
55656.51323 &689990 &18.149 $\pm$ 0.023 & UKIRT & K \\
55663.41583 & 1287023 & 18.379 $\pm$ 0.294 & UKIRT & K \\
55664.46310 & 1376628 & 18.401 $\pm$ 0.272 & UKIRT & K \\
\hline
\hline
\end{tabular}
\end{center}
\caption{Infrared photometry of \event~}
\label{ir}
\end{table}%

\subsection{Limits on rapid variability}
We obtained 81, 20-second observations of \event in the $i$-band with OSIRIS on the GTC on 30 March 2001. 
We can use these to place
limits on any rapid variability within the optical data, similar to that observed in the X-rays. Photometry of
the counterpart relative to two comparison stars in the image, shows no sign of variability over the time
baseline of the observations. Given the non-detection of any afterglow emission in the deep r-band images 
taken with Gemini-N the non-detection 
of rapid variability within these images is not surprising. 

In addition, the short timescales for IR observations naturally lend themselves to studies of short time scale variation. 
We therefore photometer our individual NIRI frames separately, and find no evidence for rapid variability during their
timeframe. 

At first sight this would seem to imply that the IR variations, while tracking the broad shape of the X-ray lightcurve, do not show evidence for
rapid variation. However, we note that at the time of our first epoch of NIRI observations (which provide 18x60s resolution) the X-ray
afterglow of \event~ was also not apparently rapidly variable.

\subsection{Optical spectroscopy}
We obtained multiple epochs of spectroscopy of \event~ using the GTC, Gemini-North and 
Keck II telescopes. Data reduction and analysis of these data is provided below. 

The first spectroscopy of \event~ was obtained using
the OSIRIS instrument on the 10.4m Gran Telescopio CANARIAS (GTC),
starting 15.7 hours after the burst. We used
the R300B grism with a 1 arcsecond wide slit and a $2 \times 2$ binning,
taking three exposures of 1200 seconds starting on 04:43 UT on March 29,
2011. Data were reduced using tasks in {\sc IRAF}, using calibration data
(bias frames, flatfields and arc frames) taken the same night. Flux
calibration was done using spectra of standard star Ross 640.

We obtained two epochs of spectroscopy of the transient and its host
galaxy using the Gemini Multi-Object Spectrograph (GMOS) on Gemini-North.
The first spectrum was taken using the R400 grating with a 1.0\arcsec\
slit width and using a standard $2\times2$ binning. Four 600\,s exposures were
taken, using dithers in spatial and dispersion directions (central
wavelengths 6000 and 6050\,\AA), with observations starting at 13:16 UT on
March 29, 2011.

The second spectrum was taken using the same grism, slitwidth and binning,
but used a central wavelength of 8000\,\AA. Two exposures of 720\,s were
obtained in the so-called ``nod \& shuffle'' mode to achieve accurate
subtraction of nightsky emission lines 
\cite{GlazebrookBlandHawthorn2001}. 
No dithering in dispersion direction was performed. Observations
started 14:51 UT on April 1st, 2011.

The data of both epochs were reduced using the Gemini GMOS data
reduction packages within IRAF. Flatfields and Cu+Ar arc lamp spectra
taken just before the science data were used for calibration. Flux
calibration was achieved using observations of standard star Hiltner 600
\cite{Hamuy1992} and photometry was used to bring the spectra to an
absolute flux scale. Finally, the spectra were corrected for a Galactic
reddening of $E(B-V) = 0.02$, assuming $R_V = 3.1$.

Finally, we obtained moderate-resolution ($R \equiv \lambda / \Delta \lambda \approx 2500$) spectra at the location of \event\ with the DEep Imaging Multi-Object Spectrograph [DEIMOS; \cite{fpk+03}] mounted on the 10\,m Keck II telescope on 2011 March 31 (beginning at 14:55 UT) and 2011 April 4 (beginning at 14:38 UT).  The instrument was configured with the 600\,lines\,mm$^{-1}$ grating with a central wavelength of 6900\,\AA\ (7200\,\AA) on 2011 March 31 (2011 April 4), providing a spectral resolution of 3.5\,\AA\ (FWHM) and wavelength coverage from $\approx 4500$--9500\,\AA.  
We implement a modified version of the DEEP2 DEIMOS pipeline\footnote{See http://astro.berkeley.edu/$\sim$cooper/deep/spec2d/} to rectify and background-subtract our spectra. The pipeline bias-corrects, flattens, traces the slit, and fits a two-dimensional wavelength solution to the slit by modeling the sky lines. This final step provides a wavelength for each pixel. The slit is then sky subtracted (in both dimensions) and rectified, producing a rectangular two-dimensional spectrum where each pixel in a given column has the same wavelength. From this point we proceed with standard reduction procedures, using the standard star BD+33 2642 for flux calibration.
 To account for slit losses, we renormalized each spectrum based on the observed $R$-band magnitude of the optical counterpart ($R = 22.6$\,mag).

\subsection{Host spectroscopic properties}

In our spectra we find
nebular emission lines corresponding to [O{\sc ii}] $\lambda \lambda$3726,3728, H$\beta$, [O{\sc iii}] $\lambda \lambda$4959,5007, H$\alpha$, [N{\sc ii}] $\lambda$6583, and [S {\sc ii}] $\lambda \lambda$6716,6731, all at a common redshift of $z = 0.3534 \pm 0.0002$, consistent with previously reported values \cite{GCN.11833,GCN.11834}.  A listing of all well-detected emission lines, with emission fluxes and centroids determined from Gaussian fits, is provided in Table~\ref{tab:emlines}, based on measurements from 
our Keck spectroscopy.  The uncertainties in the line fluxes are dominated in nearly all cases by uncertainty in the strength of the continuum emission.  All lines are unresolved, both spectrally and spatially, and show no significant evidence of variability between epochs (see Figure 2, main article).

  \begin{table}
\begin{center}
\begin{tabular}{llll}
\hline
Observed Wavelength & Line Flux & Identification & Redshift \\
\AA & ($10^{-17}$\,erg\,cm$^{-2}$\,s$^{-1}$) \\
\hline
    $5043.43 \pm 0.53$ & $3.5 \pm 0.6$ & [OII] & 0.35357 \\
    $5047.41 \pm 0.75$ & $8.6 \pm 1.4$ & [OII] & 0.35362 \\
    $6578.75 \pm 0.42$ & $2.7 \pm 0.6$ & H$\beta$ & 0.35327 \\
    $6711.26 \pm 0.39$ & $1.4 \pm 0.4$ & [OIII] & 0.35337 \\
    $6775.85 \pm 0.47$ & $4.0 \pm 0.5$ & [OIII] & 0.35332 \\
    $8881.45 \pm 0.55$ & $7.6 \pm 1.0$ & H$\alpha$ & 0.35329 \\
    $8908.00 \pm 0.83$ & $1.2 \pm 0.5$ & [NII] & 0.3509 \\
    $9089.11 \pm 0.71$ & $1.1 \pm 0.3$ & [SII] & 0.35326 \\
    $9109.13 \pm 0.69$ & $0.9 \pm 0.3$ & [SII] & 0.35335 \\
\hline
\end{tabular}
\end{center}
\caption{Optical emission lines from \event~ based on our Keck DEIMOS observations. These line fluxes are also consistent with
those inferred from our GTC and Gemini observations.}
\label{tab:emlines}
\end{table}%

We can estimate the extinction along the line of sight to the source using the observed intensity ratios of host-galaxy Balmer emission lines.  We find that $(L_{\mathrm{H}\alpha} /
L_{\mathrm{H}\beta})_{\mathrm{obs}} = 2.8 \pm 0.6$.  Assuming Case B recombination \cite{o89} and the relation from \cite{c01}, we find $E(B-V)_{\mathrm{gas}} = -0.01 \pm 0.15$ mag.  Applying the extinction law derived for star-forming galaxies [e.g., \cite{c01}], this corresponds a limit on the rest-frame $V$-band extinction for the stellar continuum of $A_{V} < 0.8$\,mag (3$\sigma$).

To determine the origin of these emission lines, we resort to a diagnostic, or BPT diagram [Baldwin, Phillips \& Terlevich \cite{bpt81,vo87})].  The atoms could be ionised by the hard power-law spectrum generated by gas accretion onto a central SMBH (i.e., an AGN), UV photons from young, massive O and B stars (i.e., star formation), or as part of a phenomenon known as a Low Ionisation Nuclear Emission-line Region [LINER; \cite{h80}], which are likely related to AGN, possibly resulting from changes to the geometry of the disc at low accretion levels (e.g., \cite{h08}).  The observed ratios between various atomic species, since they are highly sensitive to the nature of the ionizing continuum, can be used to distinguish between these alternatives.

In figure ~\ref{fig:bpt}, (SOM) we plot the ratio of $L_{\left[ \mathrm{O\,III} \right]
    \lambda 5009} / L_{\mathrm{H}\beta}$ against $L_{\left[ \mathrm{N\,II} \right] \lambda 6583} / L_{\mathrm{H}\alpha}$.  Empirical \cite{kht+03,hfs97} and theoretical \cite{kd02} dividing lines between the various classes of objects are plotted, as well as a series of analogous measurements from the MPA/JHU value-added SDSS catalog\footnote{See http://www.mpa-garching.mpg.de/SDSS.}.  The optical counterpart of \event\ clearly falls within the phase space of star-forming galaxies.  In other words, there is no evidence for nuclear activity (i.e., an AGN) based solely on the optical spectra \cite{GCN.11874}.

\begin{figure}
\centerline{
%	\centerline{\psfig{file=complin.ps,angle=270,width=9cm,clip=5mm}}
    \includegraphics[width=12.5cm, angle=0]{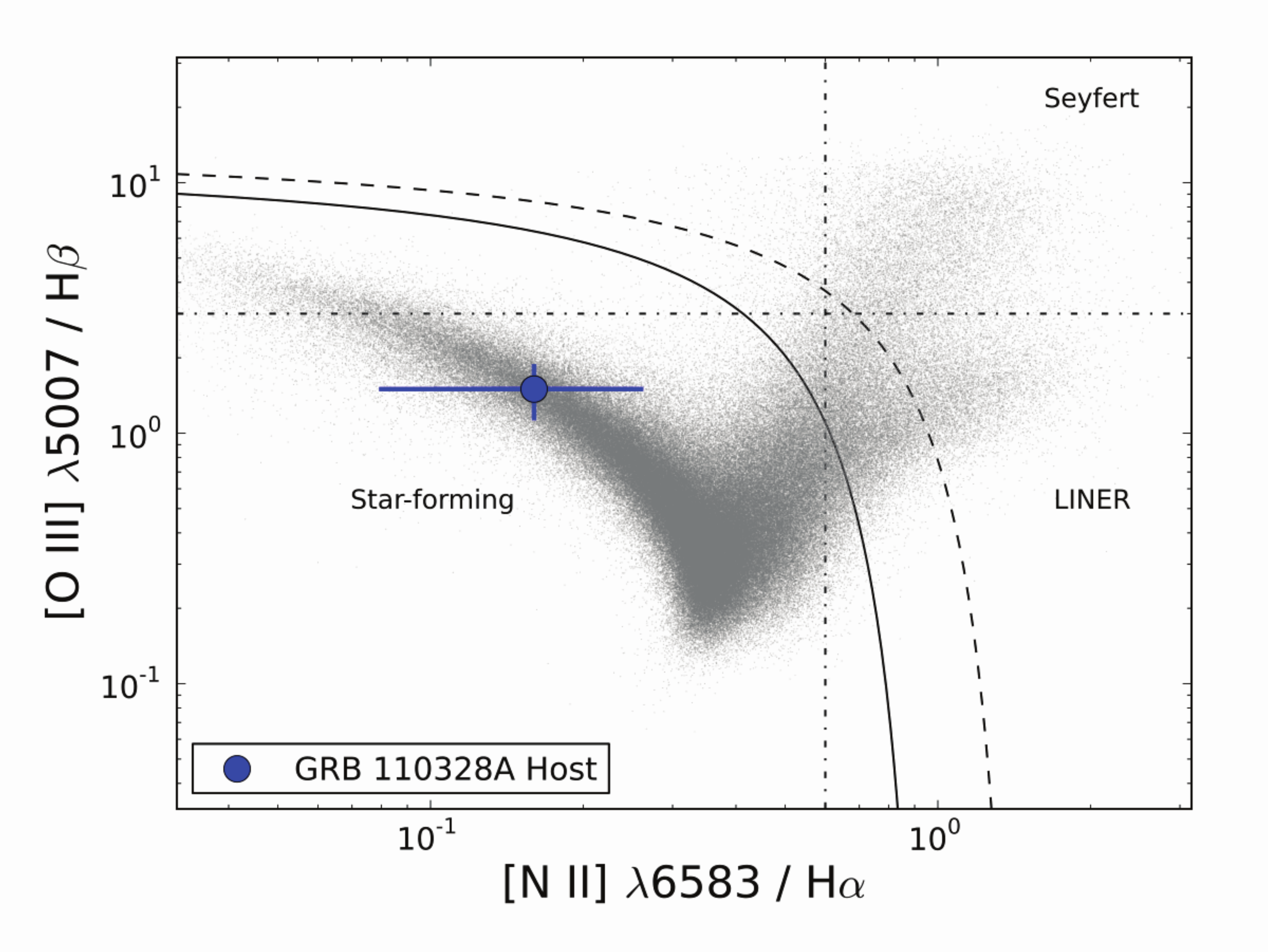}}
  \caption{Diagnostic emission-line diagrams for the optical counterpart of     
  \event.  The empirical dividing line between star-forming and active galaxies 
  from \cite{kht+03} is shown as the solid line, while analogous dividing lines   
  from \cite{hfs97} are indicated with dashed-dotted lines.  The theoretical 
  dividing line from \cite{kd02} is plotted as a dashed line.  Analogous 
  measurements for SDSS galaxies from the MPA-JHU value-added catalog are 
  shown as gray dots.  The optical counterpart falls firmly on the locus of    
  star-forming galaxies, with no indication for any previous nuclear activity (i.e., 
  an AGN).}
\label{fig:bpt}
\end{figure}

We can calculate the current star formation rate based on the strength of the H$\alpha$ and [O {\sc ii}] emission lines.  Using the calibration from \cite{k98}, we find SFR$_{H\alpha} = 0.3 \pm 0.1$\,$M_{\odot}$\,yr$^{-1}$, and SFR$_{[O II]} = 0.7 \pm 0.2$\,$M_{\odot}$\,yr$^{-1}$.  Taking the average, we adopt a value of SFR$ \approx 0.5$\,$M_{\odot}$\,yr$^{-1}$; the errors above likely underestimate the true uncertainty, as they do not incorporate any effects of the assumed star-formation calibration.

Using the derived line fluxes from Table~\ref{tab:emlines}, we have measured the metallicity of the host galaxy based on our DEIMOS
and GMOS spectra following the technique described in \cite{mkk+08}.  Using the scales from \cite{pp04} (PP04), we find $12 + \log(\mathrm{O/H}) = 8.38 \pm 0.16$ based on the \ion{N}{2} / H$\alpha$ diagnostic (PP04-N2H$\alpha$), and $12 + \log(\mathrm{O/H}) = 8.43^{+0.12}_{-0.14}$ based on the \ion{O}{3} / \ion{N}{2} (PP04-O3N2) prescription.  Likewise, we find $12 + \log(\mathrm{O/H}) = 8.60^{+0.19}_{-0.17}$ on the scale of \cite{kd02} (KD02), and $\log(\mathrm{O/H}) = 8.53 \pm 0.22$ on the scale of \cite{m91} (M91).  
Given the most recent estimate of the solar oxygen abundance [$\log(\mathrm{O/H}) = 8.70$; \cite{ags+09}], together with the relative robustness of $T_{e}$-based metallicity scale \cite{bgk+09}, which the PP04-O3N2 scale is close to, we conclude the host metallicity is $Z \approx 0.5 Z_{\odot}$.  A comparison with SDSS samples of galaxies, as well as those of SN Ib/c and GRBs is shown in Figure~\ref{fig:metallicity} (SOM). This suggests that the host of \event~ has a typical metallicity for its luminosity.

\begin{figure}
\centerline{
%	\centerline{\psfig{file=complin.ps,angle=270,width=9cm,clip=5mm}}
    \includegraphics[width=12.5cm, angle=0]{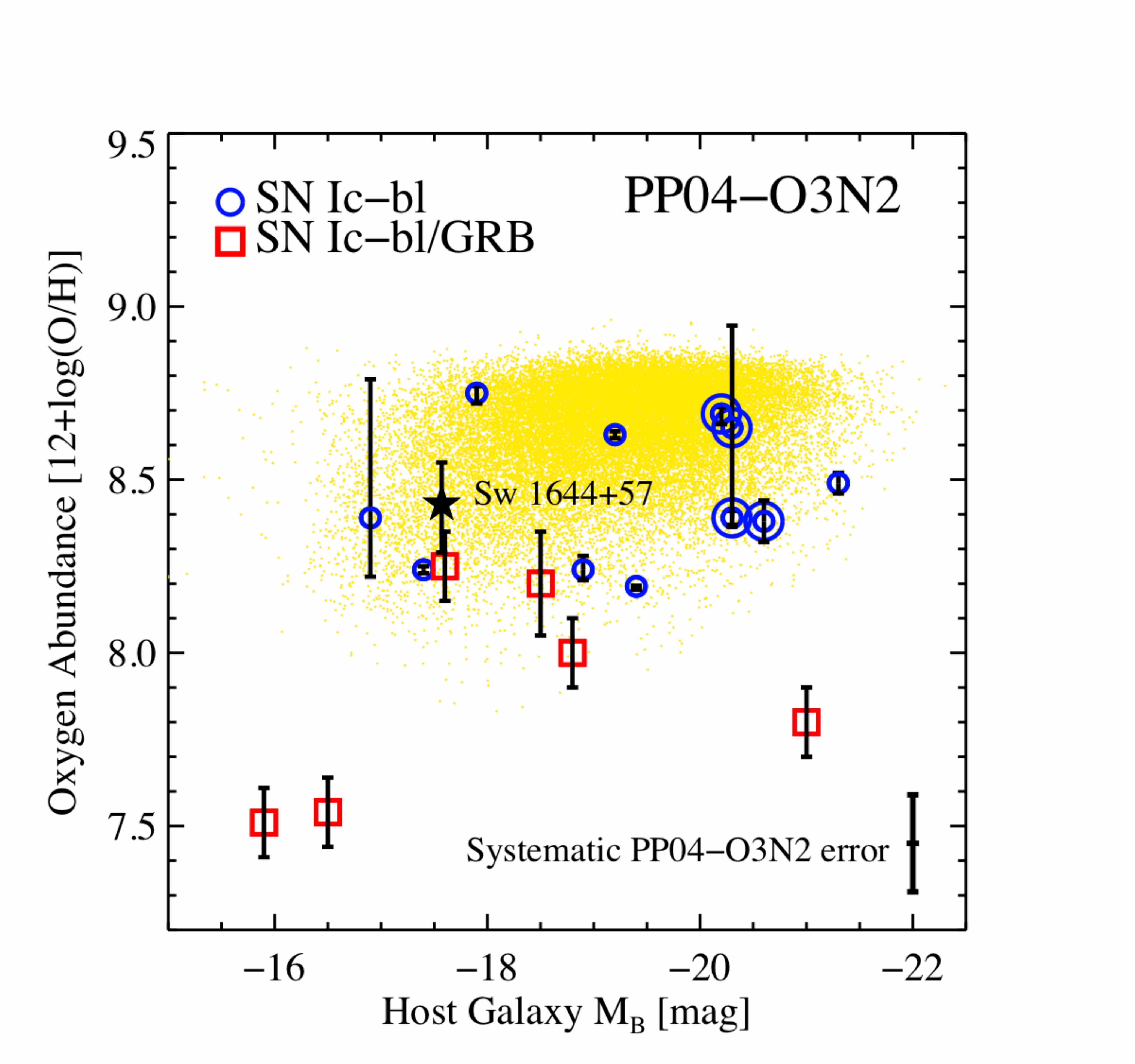}}
  \caption{Host-galaxy luminosity ($M_B$) and host-galaxy metallicity (in terms of oxygen abundance) at the sites of Swift J164449.3+573451 ("Sw 1644+57"), nearby broad-lined SN Ic (ÒSN Ic-blÓ: blue circles) and broad-lined SN Ic connected with GRBs (ÒSN-bl/GRB)Ó: red squares). The oxygen abundances are in the (20) PP04-O3N2 scale. Yellow points are values for local star-forming galaxies in SDSS (25), re-calculated in the (20) scale for consistency, and illustrate the empirical luminosity-metallicity (LÐZ) relationship for galaxies.
   The host galaxy of \event~ is broadly
    consistent with the bulk of SDSS galaxies at that luminosity. For more details see \cite{mkk+08}}
\label{fig:metallicity}
\end{figure}

In addition to emission lines, continuum emission is detected over most of the spectral range, showing
many (stellar) absorption features (e.g., G band, Na\,{\sc I}, Mg\,{\sc I},
Fe\,{\sc I}, Ca H + K, Balmer line absorption) characteristic of an older
(older than a few hundred Myr) population. To derive the H$\beta$ emission
line flux we fit simultaneously the continuum, the stellar absorption
component and the emission line flux, where the lines are fit using
Gaussians. In the first Gemini spectrum, which has highest signal to
noise, we find a restframe equivalent width for the H$\beta$ absorption of
$8.2 \pm 1.1$\,\AA. A similar fit on H$\delta$ is not possible as the
signal to noise is too low to reliably characterize the absorption
component. A similar fit on H$\gamma$ suffers from the presence of
residuals from a nearby telluric emission line in the red absorption wing,
but a $\sim4 \sigma$ detection of absorption was found with restframe equivalent
width $11.7 \pm 2.4$\,\AA.

\subsection{Host photometric properties}
In principle we can use SED fitting to the observed host photometry to
extract measurements of the luminosity of the host of \event~ in various
wavebands. Of particular importance is the stellar mass, which among the
global properties of a galaxy is most closely linked to the size of its
the central black hole.

The stellar mass of a galaxy is normally estimated using a measurement
of the rest-frame $K$-band (2 $\mu$m) luminosity multiplied by a
constant (of order unity in Solar units). Unfortunately, as long as the
transient remains bright, a direct measurement of the infrared host flux
is impossible.  Fortunately, the transient contributes much less to the
optical light: the resolved HST F606W observations constrain the
contribution of the transient to at most 30\% of the light in this
filter (see below), and blueward of this band the contribution should be
even less.  Even in the NIR, the faintest measurement in each filter can
be used to place a conservative upper limit on the host-galaxy flux;
further constraints are provided by WISE pre-imaging of the field at 3.4
$\mu$m.  Hence we can fit template models using these values as hard
upper limits to allowable IR flux.

In Figure~\ref{hostsed}, we plot the photometric constraints on the
galaxy flux (that is, with the transient light removed) against
different models of the stellar SED using the population-synthesis
templates of \cite{bc2003}. While a diverse set of different models
(from young starbursts to evolved, quiescent populations) are nominally
consistent with the observations, we can nevertheless bracket the
possible range of $K$-band fluxes and back out a maximum and minimum
luminosity.  Using a stellar mass-to-light ratio of $M_*/L_{K}$ = 0.4 in
Solar units (following \cite{cc2010}) the range of stellar masses
consistent with the data is approximately $10^9-10^{10} M_{\odot}$.

\begin{figure}[ht]
\centerline{
%	\centerline{\psfig{file=complin.ps,angle=270,width=9cm,clip=5mm}}
    \includegraphics[width=15cm, angle=0]{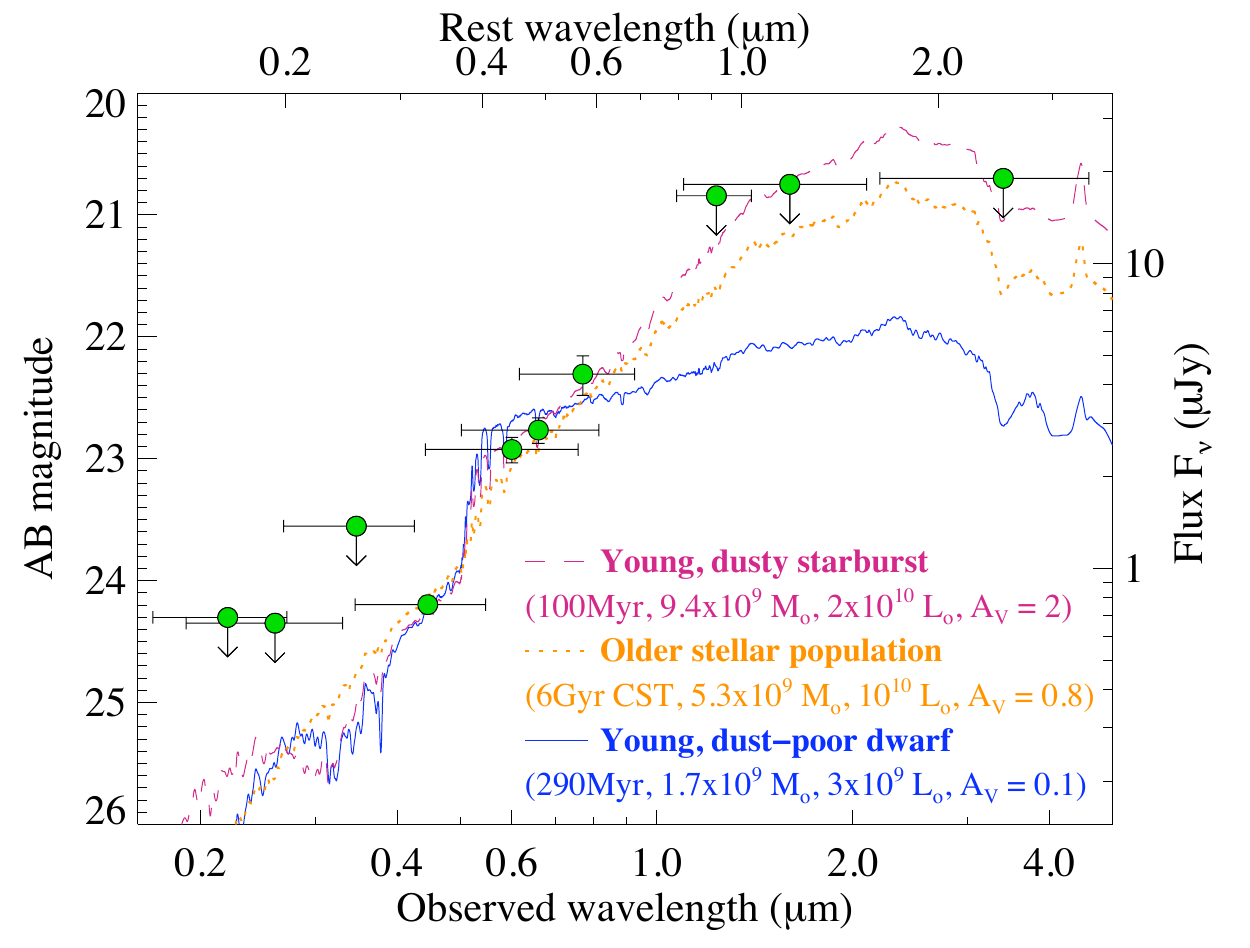}}
\caption{Photometry of the host galaxy of \event, matched to a variety of population-synthesis templates [BC2003] bracketing the range of possible IR colors.  Optical/UV upper limits are provided by the UVOT, with the optical NIR points coming from our ground-based imaging and HST.  The upper limit at 3.4 $\mu$m is from pre-imaging from WISE. Three population-synthesis bracketing the variety of possibilities are shown.  Population ages (a single stellar population is assumed for the young models; continuous star formation is assumed for the older model), inferred stellar masses, bolometric luminosities, and total extinction for each model are indicated.  (For clarity, the templates have been smoothed to a resolution of $\lambda$/$\Delta \lambda$ = 100.)}
\label{hostsed}
\end{figure}

\subsection{Hubble Space Telescope}

We obtained {\em Hubble Space Telescope (HST)} observations of \event~ on 4 April 2011, beginning at 03:03 UT. 
At this epoch we observed four dithered exposures in F110W in the IR channel of the Wide Field Camera 3, and
3 dithered exposures in the F606W within the UVIS channel. These observations were combined with 
{\tt multidrizzle} to provide images with final pixel scales of 0.067 and 0.033\arcsec\ respectively.

The {\em HST} images contain two stars that are common to both the 2MASS \cite{2MASS} and UCAC3 \cite{UCAC3} catalogs (see table \ref{ucac}).   Both of these catalogs have small astrometric errors ($ \leq 0.07$\arcsec); we therefore averaged their astrometric positions to provide an astrometric reference for our {\em HST} images.   While both of these stars are heavily saturated in the WFC3/UVIS images, the saturation is mild in the WFC3/IR images.  Thus centroiding the light will still give a position to an accuracy better than than our drizzled pixel scale of $0.067$\,arcsec.   Additionally it is possible to use the diffraction spikes of a bright star to estimate its location.   These two methods agreed, again to a fraction of a drizzled IR pixel.    As expected, the orientation and scale of the HST IR image World Coordinate System agreed within the errors with
the astrometric positions of the two stars.   We therefore used the stars only to update the absolute
astrometric reference position of the image and not its orientation or scale.   Following
this procedure, we obtain a
position for the IR source of RA = 16:44:49.9345, Dec = +57:34:59.673 J2000, with an estimated
error of $0.07$\arcsec\ in each coordinate.  Our derived position differs from the VLBA (see below)
position by less than $0.03$\arcsec\ in both coordinates.

The host galaxy of GRB~110328A is clearly resolved in the {\em HST} WFC3 F606W image.   
The galaxy profile has a full-width half maximum (FWHM) of about $0.13$\arcsec\ compared to 
a FWHM of  about $0.085$\arcsec\ for stars in the image.   To determine a maximum 
contribution to the host galaxy magnitude from the transient we subtracted scaled stellar point source functions (PSFs) from both the centroid of the galaxy
light, as well as from location of the IR source, which in our best estimate lies about $0.016$\arcsec, 
from the
galaxy centroid.    We find that we subtract up to 20\% of galaxy light in this manner without causing unreasonable deviations in the galaxy's light profile.   
However, a PSF containing $ \sim 30\% $ of the
galaxy's light causes clear errors, with depression of the central galaxy flux relative to the surroundings,
and thus is a fairly hard upper limit to the total light in the transient.
This corresponds to an F606W magnitude of 24.1 AB.

\begin{figure}[ht]
\centerline{
%	\centerline{\psfig{file=complin.ps,angle=270,width=9cm,clip=5mm}}
    \includegraphics[width=15cm, angle=0]{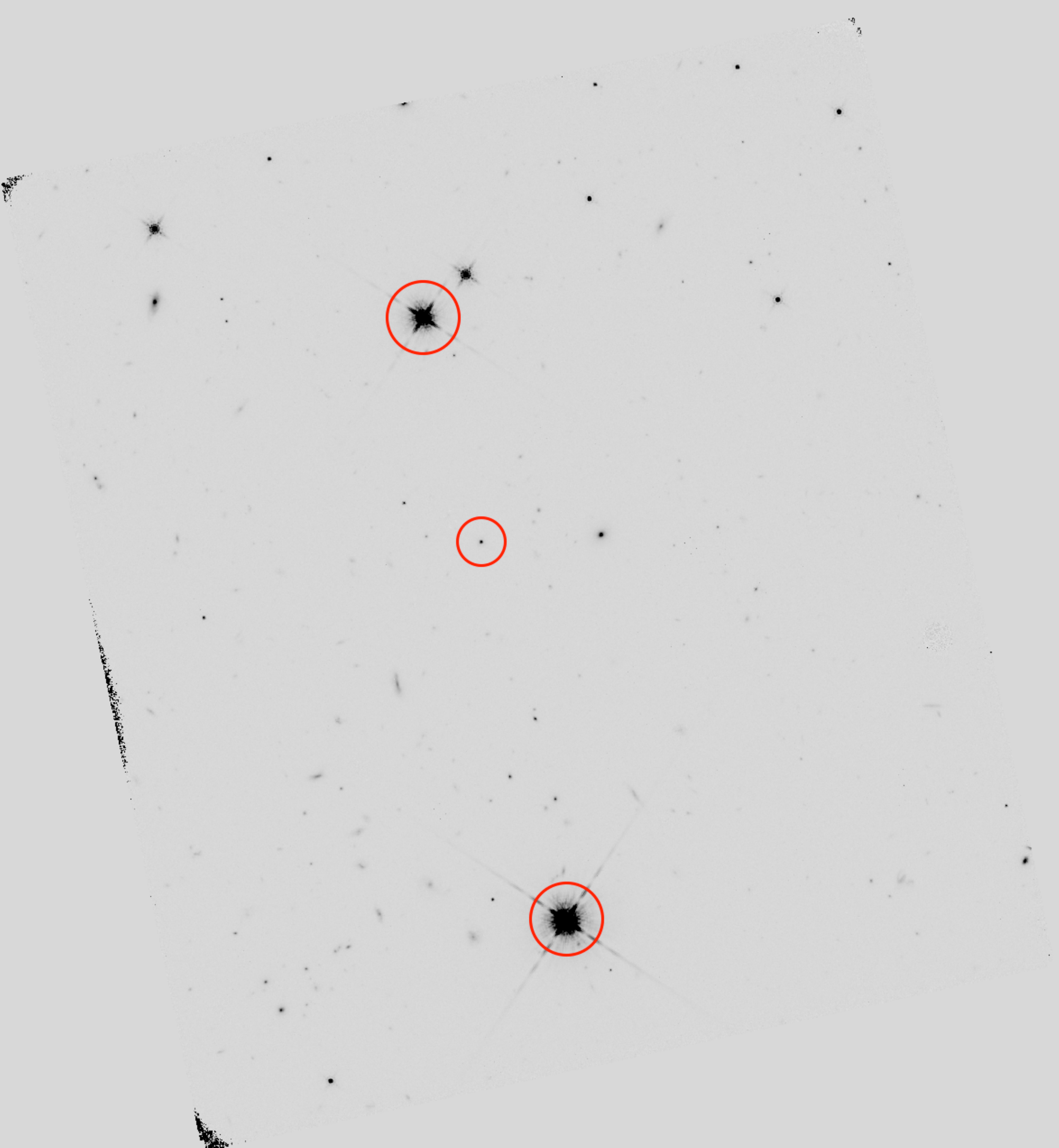}}
\caption{A wide field view of our WFC3 IR channel observations of \event. The image shows the full field of view, and marks the location of two 2MASS and UCAC3 stars
used for astrometric calibration (A and B in table~\ref{ucac}, as well as the central position of the host galaxy of \event. }
\label{nirfov}
\end{figure}

\begin{table}
\begin{center}
\begin{tabular}{cccc}
\hline
Star  & ID                   & RA (J2000)     & Dec (J2000)\\
\hline
A& 2MASS 16445096+5735316 & 16:44:50.964
& +57:35:31.64 \\
B & 2MASS 16444843+5734057 & 16:44:48.432
& +57:34:05.74 \\

A& 3UC 296-121315         & 16:44:50.965   & +57:35:31.69 \\
B &3UC 296-121311         & 16:44:48.432   & +57:34:05.82 \\
\hline
\end{tabular}
\caption{2MASS and UCAC3 positions for two stars lying the field of view of our WFC3 UVIS and IR observations, and used
for astrometric calibration. The ID's A and B refer to the locations of the stars in Figure~\ref{nirfov} }
\label{ucac}
\end{center}
\end{table}

\subsection{Chandra X-ray Observatory (CXO)}

Following the discovery of the host galaxy of \event~ we initiated Target of Opportunity observations with the {\em Chandra} X-ray Telescope. 
These observations were obtained of 4 April 2011, beginning at 02:30 UT, and utilised the High Resolution Camera in imaging mode
(HRC-I). Using the standard cleaned event file we extract both images and lightcurves of the data, with the 
lightcurve shown in Figure~2 of the main paper, in 100s bins. The afterglow of \event~ is strongly
detected in the images, with a total of $\sim 50,000$ counts obtained over the course of the observation, at a mean count rate of $\sim 3$ counts
per second. Since the HRC is primarily an imaging and timing instrument we do not create spectra, but instead 
convert counts to flux by assuming the
spectral model from {\em Swift} XRT PC mode observations taken at approximately the same epoch. 

We additionally perform astrometry between the CXO observations and our ground based Gemini and UKIRT observations. 
We identify 5 X-ray sources with optical counterparts in our UKIRT/Gemini image, and these are shown in Table~\ref{CXO}. We
then perform relative astrometry between these images, allowing us to place the position of the X-ray afterglow on the ground
based frames with an accuracy of $\sim 0.25$\arcsec. The resulting position is offset $0.07 \pm  0.25$ \arcsec from
the centre of the host galaxy, fully consistent with a nuclear origin. 

\begin{table}[htdp]
\begin{center}
\begin{tabular}{lll}
\hline
Object & RA(J2000) & DEC(J2000) \\ %& counts \\
\hline
Afterglow   &      16:44:49.905  &  57:34:59.82     \\   
1                 &      16:45:05.036    & 57:37:41.41      \\  
2                   &    16:44:22.354    & 57:35:43.88       \\ 
3                     &  16:44:21.893    & 57:36:15.17        \\   
4                    &   16:44:41.285    & 57:35:04.94         \\    
5                      & 16:45:02.019    & 57:37:40.99          \\     
\hline
\end{tabular}
\end{center}
\caption{The locations of Chandra sources used for optical/X-ray astrometry. The positions are given in the world co-ordinate system of the Chandra image, but
since relative astrometry is performed, the precise co-ordinates are not important.}
\label{CXO}
\end{table}%

%\subsection{Herschel}

\subsection{Swift Observations} 
\subsubsection{XRT}
In Figure~1 (main article) we plot the X-ray lightcurve of \event\ from the {\em Swift}-XRT and {\em Chandra}. This lightcurve was
created following the method described in \cite{evans09}, and retrieved from the online XRT repository \footnote{http://www.swift.ac.uk/xrt\_curves}. As noted in the main text the lightcurve is characterized by large scale variability, even at relatively late times
after burst the variability is of order a factor $10^2$, on timescales of 1000 seconds. We highlight this in figure~\ref{fold} (SOM), where
we show zoomed in regions around two flares, which occurred on the second day. In particular we have overlayed these flares
to a common time axis (by folding with a time period of 34300s) to demonstrate their broad similarity in morphology, especially during the decay.

\begin{figure}[ht]
\centerline{
%	\centerline{\psfig{file=complin.ps,angle=270,width=9cm,clip=5mm}}
    \includegraphics[width=15cm, angle=0]{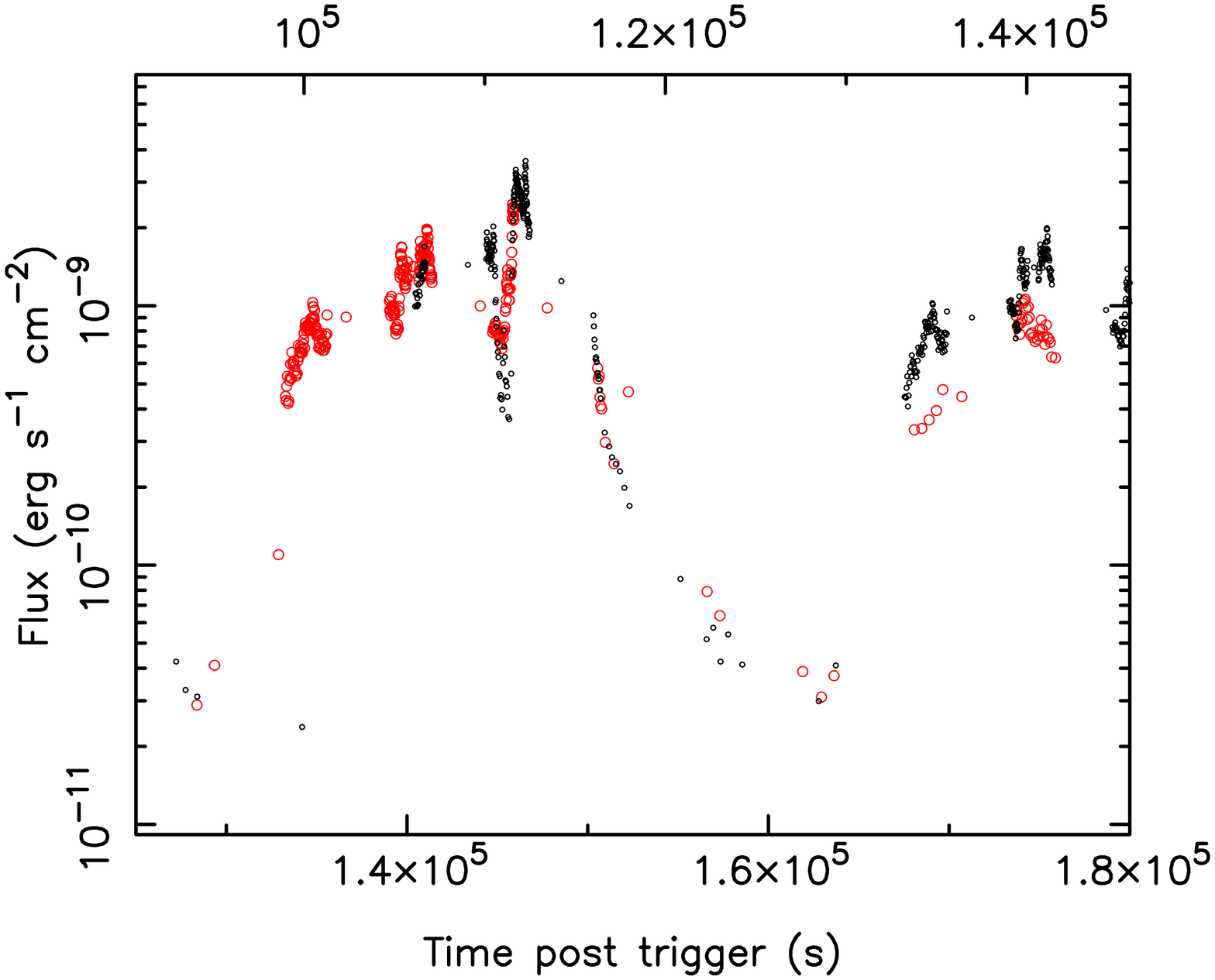}}
\caption{Zoom in of the X-ray light curves of the two bright flares on the second
day post-trigger.  The earlier flare is plotted in black points and corresponds
to the upper time axis, and the later flare in red corresponds to the lower
time axis.  The fold period is 34300s. Note that there has been no flux normalization applied, such that
the peak luminosity of the flares is very similar.  }
\end{figure}

In order to fit and describe the strong observed spectral
variability \cite{gcn11846,atel3242},
we divide the XRT spectra into time-contiguous segments, containing
$>500$ cts (0.3--10 keV).  This results in 139 PC mode spectra
spanning the time 6.08 ksec to 1372.09 ksec after the BAT trigger
and 461 WT mode spectra spanning the time 1.49 ksec to 191.67 ksec
after the BAT trigger (with corresponding total exposures
of 228.77 ksec and 14.33 ksec, respectively).  The spectral reductions
and fitting are performed as described in \cite{bk07}.

We find that the time-resolved spectra are well-fit by an absorbed powerlaw model with time-varying
normalization, absorption column $N_H$, and photon index $\Gamma$.
The average column in addition to the Galactic value ($1.66 \times
10^{20}$ cm$^{-2}$; \cite{kalberla05}) is $(1.2\pm0.1) \times
10^{22}$ at $z=0.35343$, and time variations about this value do not appear to
be statistically significant.  However, the flux and photon index exhibit
a strong anti-correlation (see below).
The time-average photon index is $\Gamma=1.80+/-0.25$, dominated by
the PC mode data with an average flux of
$(1.6\pm0.9) \times 10^{-10}$ erg cm$^{-2}$ s$^{-1}$.  (The quoted
flux uncertainty here reflects the flux variation rather than the
statistical uncertainty in any given 500 count epoch.)  The WT mode
data span a higher flux region of the emission ($(2.3\pm1.4) \times
10^{-9}$ erg cm$^{-2}$ s$^{-1}$) and have a harder mean photon index
of $\Gamma
= 1.6 \pm 0.2$, which is consistent with that measured for the BAT
at similar epochs (see below).  We note that the
BAT lightcurve closely tracks the X-ray lightcurve.

As discussed in Kennea et al. \cite{atel3242} and Bloom et al. \cite{gcn11846},
the X-ray hardness tracks the X-ray flux.
Consistently, we see that the best-fit powerlaw index increases with
decreasing flux, with high-flux values corresponding to the quoted value
of 1.6 for the WT mode data above and low-flux values just after the
bright WT mode X-ray flares of $\Gamma=3$.  The spectra in these
regions of flux decline can also be well-modelled assuming a (nearly)
constant X-ray flux normalization and a fixed, single powerlaw index
(e.g., $\Gamma=1.6$), but allowing for an exponential cutoff which starts
above the X-ray bandpass and passes into the X-ray bandpass as the source
flux declines, reaching $E_{\rm cut} \approx 1$ keV.  The specific values
for these cutoffs depends on the value we fix for $\Gamma$.

\subsubsection{BAT}
In Figure 1 (main article) we also plot the BAT data. These data are taken
from the online repository\footnote{http://swift.gsfc.nasa.gov/docs/swift/results/transients/}
and were subsequently re-binned to have S/N$=3$ (15-50 keV band). To convert these count
rates to fluxes, we jointly fit the overlapping XRT data during the flare region from 2342.8 to 2715.1 ksec.
Consistent with BAT only fits from \cite{gcn11842} and the XRT-only fits described above, we find the BAT and XRT spectra are jointly
described-well ($\chi^2/\nu = 587.0253/457$) by an absorbed powerlaw spectrum with photon index $\Gamma$ =  1.62 $\pm$ 0.05.
The normalization at 1 keV is $0.48\pm0.03\pm 0.5$ mJy, and the inferred column density is
N$_H=1.50\pm 0.08 \times 10^{22}$ cm$^{-2}$ ($z=0.3434$).

There is modest evidence ($3.2-\sigma$ significant$; \Delta \chi^2=9.93$ for 1 additional degree of freedom) for a cutoff in the BAT spectrum
at $\nu F{\nu}$ peak energy of E$_p$ $71^{+66}_{-22}$ keV.  Using this spectral fit, we the convert the measured BAT count rate
to 15-50 keV fluxes, which are overplotted on in Figure 1 of the main journal. This fit does not therefore take into account the fine
detail that may be present in the spectral evolution over time, but shows broadly that the BAT flux does trace that seen with the
XRT.

\subsubsection{UVOT}

The \emph{Swift}-UVOT instrument began observing the field of \event\
1482\,s after the initial BAT trigger, starting the usual automatic
sequence of observations.  No optical or ultraviolet counterpart was
detected at the location of \event\ in either single frames or
coadded images \cite{GCN.11910}.

After the first two days, observations were obtained in ``filter of
the day'' mode.  We coadded these late-time observations using 
the {\tt HEASOFT v6.10} distribution of the UVOT reduction tools. 
In particular a single combined frame has been obtained using 
the {\tt uvotimsum} task.  Upper limits were then obtained with 
the {\tt uvotsource} task, following the procedure described by 
\cite{Poole:2008yq} and applying the latest UV filters zero points 
\cite{Breeveld:2011kx}.  In Table~\ref{uvot} we report 3$\sigma$ 
limits for the bluest UVOT filters.

\begin{table}[htdp]
\begin{center}
\begin{tabular}{llllll}
\hline
Time (UT) & Exposure (s) & Filter & Frequency ($Hz$) & Mag & Flux (mJy) \\
\hline
March 31 & 11220 &$white$&$8.64\times10^{14}$&$>23.94$&$<5.13\times10^{-4}$\\
March 31 & 11411 &$u$&$8.56\times10^{14}$&$>23.00$&$<9.05\times10^{-4}$\\
April 1 & 20899 &$uvw1$&$1.16\times10^{15}$&$>23.27$&$<4.37\times10^{-4}$\\
April 3 & 15235 &$uvm2$&$1.35\times10^{15}$&$>23.05$&$<4.65\times10^{-4}$\\
April 6 & 13030 &$uvw2$&$1.48\times10^{15}$&$>23.26$&$<3.66\times10^{-4}$\\
\hline
\end{tabular}
\end{center}
\caption{Log of {\em Swift} UVOT observations and upper limits}
\label{uvot}
\end{table}%

\subsection{VLBA Observations}

\event~ was observed on 1 and 3 April 2011 with the 
Very Long Baseline Array (VLBA).  On both dates, observations began at 0530 UT 
and continued for 4 hours.  Observations were obtained at a sky frequency of 8.4 GHz
and with a recording bandwidth of 512 Mbps and correlated in full Stokes mode.
The bright, compact source 3C 345 was used to calibrate instrumental delay and phase 
terms.  The ICRS reference source J1638+5720 was used as phase calibrator in
a switching cycle of 4 minutes with \event.  Observations
were also obtained of the nearby compact object J1657+5705 every 30 minutes.
A prior amplitude corrections were applied.  Calibration 
of the linear polarization leakage terms was obtained using J1638+5720.
Each source was imaged in Stokes I, Q, and U on both days.  We also merged
together visibility data from both dates and produced joint images.
We show an image of the source in Stokes I and in polarized intensity in Fig.~\ref{fig:vlba}.

\begin{figure}[p!]
\centerline{
%	\centerline{\psfig{file=complin.ps,angle=270,width=9cm,clip=5mm}}
    \includegraphics[width=12.5cm, angle=270]{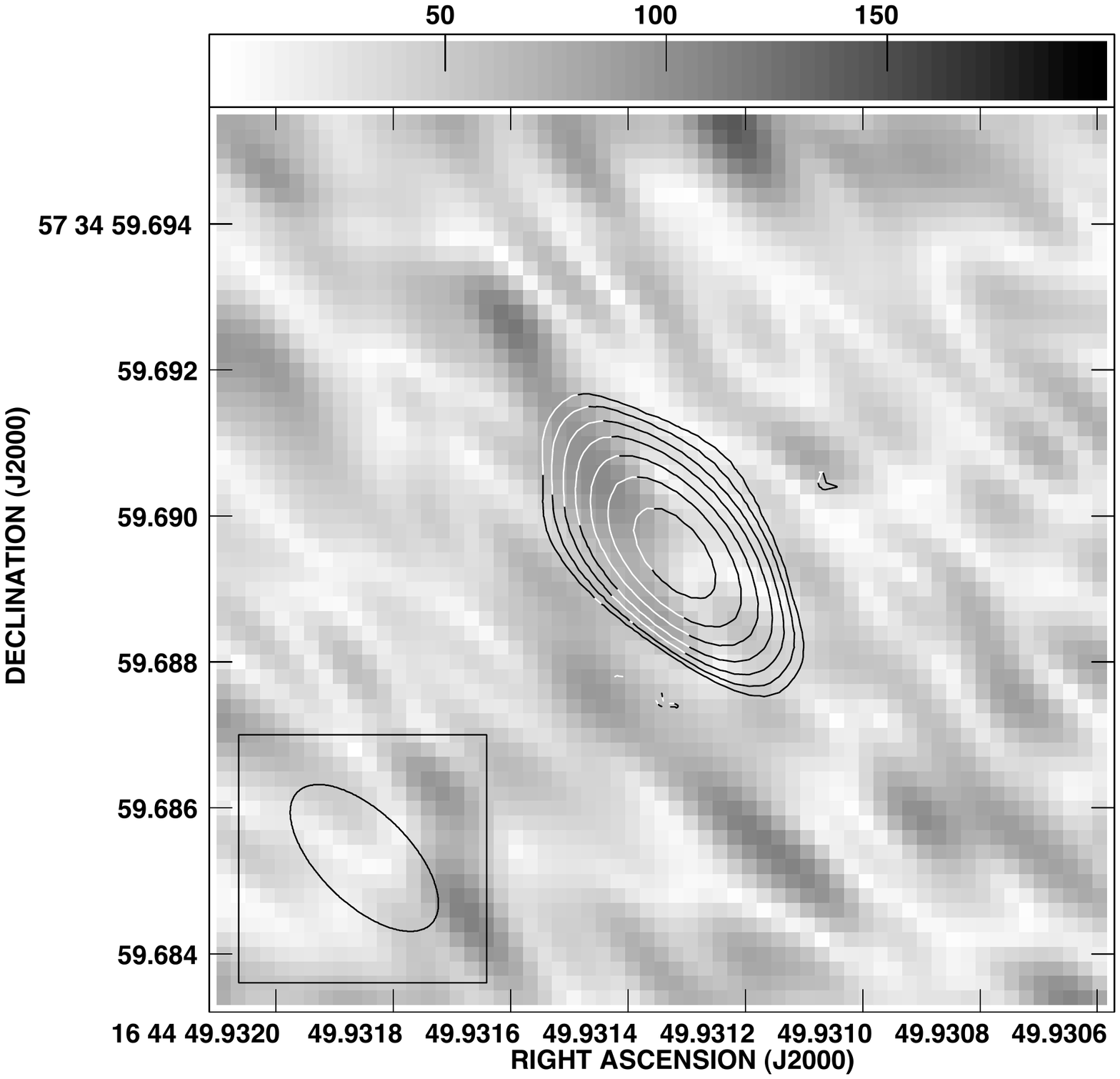}}
\caption{Image of GRB 110328A obtained with the VLBA from 1 and 3 April 2011.  
The contour levels indicate
the Stokes I image, while the gray scale indicates the polarized intensity.
Contour levels are $-$4, 4, 5.6, 8, 11.2, 16, 22.4, and 32 times the rms noise of 
37 microJy.  The synthesized beam is shown in the lower left.
\label{fig:vlba}}
\end{figure}

In Table~\ref{tab:vlba} we summarize the results for GRB 110328A.  We include
results for each date and for the merged data.  Columns give (1) the epoch
of observation; (2) synthesized beam size; (3) seconds of right ascension;
(4) arcseconds of declination; (5) flux density; (6) polarization fraction ($2\sigma$ upper limit);
and (7) rms fractional variations on hourly timescales ($2\sigma$ upper limit).  The coordinates are given relative 
to the position of (16$^h$44$^m$, +57$^\circ$34').

\begin{table}[p!]
\scriptsize
\caption{VLBA Results for GRB 110328A \label{tab:vlba}}
\begin{tabular}{lrrrrrr}
Date & Beam & $\alpha$ & $\delta$ & $S$ & $p$ & $m$ \\
     & (arcsec $\times$ arcsec) & (sec) & (arcsec) & (mJy) &  &  \\
\hline
1 April 2011 & $2.85 \times 1.76$, 48$^\circ$ & $49.9313356 \pm 0.000004360$  & $59.689571 \pm 0.00003562$ & $1.7 \pm 0.1$ & $<4.5\%$ & $<34\%$ \\
3 April 2011 & $2.37 \times 1.00$, 45$^\circ$ & $49.9313136 \pm 0.000003396$  & $59.689498 \pm 0.00003083$ & $1.7 \pm 0.1$ & $<4.7\%$ & $<34\%$ \\
\hline
Average & $2.57 \times 1.24$, 45$^\circ$ & $49.9313218 \pm 0.000002646$ & $59.689530 \pm 0.00002331$ & $1.68 \pm 0.070$ & $<2.7\%$ & \dots \\
\end{tabular}
\end{table}

The source is point-like in all three images, implying an upper limit to the size of $\sim 1$ mas.
The compact source size confirms the non-thermal nature of the source and can be used to argue
for relativistic motion \cite{bloom2011}.
The source is constant in flux density between the two epochs and there is no evidence for 
variability on shorter timescales at a level of 17\%.  No linear polarization is detected with
a $2\sigma$ upper limit of 2.7\% in the merged data.

The positions are relative to the ICRS position of J1638+5720,  
16$^h$38$^m$13.$^s$4563  57$^\circ$20'23."979.
Formally, the position of the source changes by $177 \pm 45$ microarcsec and $73 \pm 47$ microarcsec
between the two epochs.  These differences are likely dominated by systematic errors, thus we do not
believe the data require significant proper motion.  
Refined analysis including use of
updated Earth orientation parameters may permit us to reduce systematic errors, and
several further epochs are planned over the coming months. For reference note than an 
apparent motion of
$10c$ will be produce an offset of 174 microarcsec in 100 days.

\subsection{Westerbork Synthesis Radio Telescope}

We additionally obtained observations with the Westerbork Synthesis Radio Telescope (WSRT) at 1.4 and 4.8 GHz. We used the Multi Frequency Front Ends \cite{tan1991} in combination with the IVC+DZB back end in continuum mode, with a bandwidth of 8x20 MHz. Gain and phase calibrations were performed with the calibrators 3C~48 and 3C~286. The observations have been analysed using the Multichannel Image Reconstruction Image Analysis and Display (MIRIAD) \cite{sault1995} software package. We observed the source at three epochs, the first observation was carried out at 4.8 GHz, and the other two at both 1.4 and 4.8 GHz. A log of these observations and the resulting fluxes are shown in Table~\ref{radio}.

\begin{table}[htdp]
\begin{center}
\begin{tabular}{llll}
\hline
Mid-Time & Interval \& duration & Frequency & Flux Density ($\mu$Jy) \\
\hline
April 1.15 & (March 31.904 - April 1.393; 12h int.)  &    4.8 GHz  &    990 $\pm$ 29 \\
April 4.14  & (April 3.897 - 4.375; 5.3h int.)   &   4.8 GHz    & 1573 $\pm$  36 \\
April 4.16  & (April 3.920 - 4.395; 5.3h int.)    & 1.4 GHz    & 221 $\pm$ 82  \\
April 10.12 &  (April 9.880 - 10.359; 5.3h int.)  &   4.8 GHz  &    2185 $\pm$ 39 \\
April 10.14  & (April 9.903 - 10.379; 5.3h int.)    &  1.4 GHz   & 284 $\pm$ 91  \\
\hline
\end{tabular}
\end{center}
\caption{A log of observations obtained at the WSRT, showing the times of the observations, their duration, band and measured flux density.}
\label{radio}
\end{table}%

\subsection{IRAM}
The IRAM Plateau de Bure Interferometer [PdBI, France \cite{iram}] observed the Sw
1644+57 source on  31 March 
(22:10 UT  - 1 Apr 00:35 UT) at 102.5 GHz, using its compact 5-antenna configuration.
The counterpart was detected on the phase centere
co-ordinates within the 3.36 x 2.30 arcsec (98.44 deg) primary beam of the antennas. 
The flux density measurement yielded 20.8 $\pm$ 0.1 mJy.

\section*{Acknowledgements}

A.J.L. and N.R.T. acknowledge support from STFC. Swift, launched in November 2004, is a NASA mission in partnership with the Italian Space Agency and the UK Space Agency. Swift is managed by NASA Goddard. Penn State University controls science and flight operations from the Mission Operations Center in University Park, Pennsylvania. Los Alamos National Laboratory provides gamma-ray imaging analysis.
S.B.C.~acknowledges generous support from Gary and Cynthia Bengier, the Richard and Rhoda Goldman Fund, NASA/Swift grant NNX10AI21G, NASA/Fermi grant NNX1OA057G, and NSF grant AST-0908886.
A.J.vdH. was supported by NASA grant NNH07ZDA001-GLAST.
GL is supported by a grant from the Carlsberg foundation. M.M. is supported by the Hubble Fellowship grant HST-HF-51277.01-A, awarded by STScI, which is operated by AURA under NASA contract NAS5-26555.
The Dark Cosmology
Centre is funded by the DNRF. 
This work makes use of data obtained by the Chandra X-ray Observatory (OBSID = 12920). Based on observations made with the NASA/ESA Hubble Space Telescope (program ID 12447), obtained from the data archive at the Space Telescope Institute. STScI is operated by the association of Universities for Research in Astronomy, Inc. under the NASA contract  NAS 5-26555.
Based on observations obtained at the Gemini Observatory, which is operated by the 
Association of Universities for Research in Astronomy, Inc., under a cooperative agreement 
with the NSF on behalf of the Gemini partnership: the National Science Foundation (United 
States), the Science and Technology Facilities Council (United Kingdom), the 
National Research Council (Canada), CONICYT (Chile), the Australian Research Council (Australia), 
Minist\'{e}rio da Ci\^{e}ncia e Tecnologia (Brazil) 
and Ministerio de Ciencia, Tecnolog\'{i}a e Innovaci\'{o}n Productiva (Argentina). 
The United Kingdom Infrared Telescope is operated by the Joint Astronomy Centre on behalf of the Science and Technology Facilities Council of the U.K. 
UKIRT data were processed by the Cambridge Astronomical Survey Unit.
Based on observations made with the Nordic Optical Telescope, operated
on the island of La Palma jointly by Denmark, Finland, Iceland,
Norway, and Sweden, in the Spanish Observatorio del Roque de los
Muchachos of the Instituto de Astrofisica de Canarias.  
Based on observations obtained with the Samuel Oschin Telescope at the
Palomar Observatory as part of the Palomar Transient Factory project. The
National Energy Research Scientific Computing Center, which is supported by
the Office of Science of the U.S. Department of Energy under Contract No.
DE-AC02-05CH11231, provided staff, computational resources and data storage
for this project. 
The WSRT is operated by ASTRON (Netherlands Institute for Radio
Astronomy) with support from the Netherlands foundation for Scientific
Research.
Some of the data presented herein were obtained at the W.M. Keck Observatory, which is operated as a scientific partnership among the California Institute of Technology, the University of California and the National Aeronautics and Space Administration. The Observatory was made possible by the generous financial support of the W.M. Keck Foundation. We acknowledge support by the
Spanish Ministry of Science and Innovation (MICINN) under the project
grant AYA 14000-C03-01 (including Feder funds)" 
The NOT and the Gran Telescopio Canarias (GTC) are
installed in the Spanish Observatorio del Roque de los Muchachos of the
Instituto de Astrofisica de Canarias in the island of La Palma.

}
\end{document}